\documentclass[aps,prd,twocolumn,nofootinbib]{revtex4}
\usepackage{graphicx}
\usepackage{hyperref}
\usepackage{slashed}
\usepackage{color}
\usepackage{subfigure}
\usepackage{ulem}

\begin{document}

\title {Production of excited heavy quarkonia in \\$e^+e^-\to \gamma^*/Z^0 \to |(Q\bar{Q})[n]\rangle +\gamma$ at super $Z$ factory}

\author{Qi-Li Liao}
\email{xiaosueer@163.com}
\affiliation{Chongqing College of Mobile Telecommunications, Chongqing 401520, China}

\author{Jun Jiang}
\email{jiangjun87@sdu.edu.cn}
\affiliation{School of Physics, Shandong University, Jinan 250100, Shandong, China}

\author{Peng-Cheng Lu}
\email{pclu@sdu.edu.cn}
\affiliation{School of Physics, Shandong University, Jinan 250100, Shandong, China}

\author{Gu Chen}
\email{speecgu@gzhu.edu.cn}
\affiliation{School of Physics and Electronic Engineering, Guangzhou University, Guangzhou 510006, China}

\date{\today}

\begin{abstract}
Within the nonrelativistic quantum chromodynamics framework, we make a comprehensive study on the exclusive production of excited charmonium and bottomonium in $e^+e^-\to \gamma^*/Z^0 \to|(Q\bar{Q})[n]\rangle +\gamma$ ($Q=c$ or $b$ quarks) at future $Z$ factory, where the $[n]$ represents the color-singlet $n^1S_0,~n^3S_1,~n^1P_0$ and $n^3P_J$ ($n=1,2,3,4; J=0,1,2$) Fock states. 
The ``improved trace technology" is adopted to derive the analytic expressions at the amplitude level, which is useful for calculating the complicated $nP$-wave channels. 
Total cross sections, differential distributions, and uncertainties are discussed in system.
According to our study, production rates of heavy quarkonia of high excited Fock states are considerable at future $Z$ factory. 
The cross sections of charmonium for $2S$, $3S$, $4S$, $1P$, $2P$, $3P$ and $4P$-wave states are about $53.5\%$, $30.4\%$, $23.7\%$, $13.7\%$, $6.8\%$, $9.2\%$, and $9.2\%$ of that of the $1S$ state, respectively. 
And cross sections of bottomonium for $2S$, $3S$, $4S$, $1P$, $2P$, $3P$ and $4P$-wave states are about $39.3\%$, $12.3\%$, $14.3\%$, $7.1\%$, $3.1\%$, $2.7\%$, and $3.1\%$ of that of the $1S$ state, respectively.
The main uncertainties come from the radial wave functions at the origin and their derivatives at the origin under different potential models.
Then, such super $Z$ factory should be a good platform to study the properties of the high excited charmonium and bottomonium states.

\end{abstract}

\maketitle

\section{Introduction}
In comparison to the hadronic colliders like Large Hadron Collider (LHC), an electron-positron collider has some advantages, as it provides a cleaner hadronic background and the collision energy and polarization of incoming electron and positron beams can be well controlled. 
A super $Z$ factory running at the energy of the $Z^0$-boson mass with high luminosity ${\cal L}\approx 10^{34\sim36}cm^{-2}s^{-1}$ has been proposed \cite{jz}, which is similar to the GigaZ mode at an Electron-Positron Linear Collider \cite{ECFADESYLCPhysicsWorkingGroup:2001igx} and the Circular Electron-Positron Collider (CEPC) \cite{CEPCStudyGroup:2018ghi}. 
Due to the high yields of $Z^0$ bosons up to $7\times 10^{11}$ at CEPC \cite{CEPCStudyGroup:2018ghi}, it can be used for studying the production of heavy quarkonium through $Z^0$ decays.

The heavy quarkonium provides an ideal platform to investigate the properties of bound states, which is a multiscale problem for probing quantum chromodynamics (QCD) theory at all energy regions. Lots of data for the production of heavy quarkonium in different collisions are collected. Taking $J/\psi$ as an example, the cross section of the inclusive production in $e^+e^- \to J/\psi+X$ is measured by the Bell experiment \cite{Belle:2009bxr}, the two-photon scattering in $e^+e^- \to e^+e^- J/\psi+X$ is studied by the DELPHI experiment at LEP II \cite{DELPHI:2003hen}, the photoproduction in $ep \to J/\psi+X$ is explored by Zeus and H1 experiments at HERA \cite{ZEUS:2002src,H1:2010udv}, the hadroproduction in $p\bar{p} \to J/\psi+X$ is studied by CDF experiment at Tevatron \cite{CDF:2004jtw}, and the hadroproduction in $pp \to J/\psi+X$ is widely explored by ATLAS, CMS, ALICE and LHCb experiments at LHC \cite{ATLAS:2015zdw,CMS:2017dju,ALICE:2012vpz,LHCb:2013itw}. 
Meanwhile, lots of theoretical and phenomenal efforts have been made to explain the measurements and to explore QCD. We refer the readers to some review papers to get detailed  information on the status, puzzles and prospects on heavy quarkonium \cite{Brambilla:2010cs,Chung:2018lyq,Chen:2021tmf}. 

Considering the fact of the nonrelativistic nature of heavy quark and antiquark inside the quarkonium, the nonrelativistic QCD (NRQCD)  \cite{nrqcd1,nrqcd2} could be a powerful tool to study the production and decay mechanism of heavy quarkonium. 
In NRQCD framework, the relativistic effect with orders of $v_Q$ ($v_Q \ll 1$) has been separated from the nonrelativistic contributions, with $v_Q$ being the typical relative velocity between heavy quark and antiquark in the quarkonium rest frame.
$v_c^2 \approx 0.3$ for charmonium and $v_b^2 \approx 0.1$ for bottomonium.
Meanwhile, it divides the calculation into short-distance coefficients and the long-distance matrix elements. 
The short-distance coefficients describe the hard scattering of partons and can be calculated perturbatively via Feynman diagrams. 
The long-distance matrix elements describe the hadronization of Fock states with $J^{PC}$ quantum numbers into heavy quaronium and are nonperturbative parameters.

It is known that analytical expressions for the usual squared amplitudes in short-distance coefficients become complicated and lengthy for massive particles in final states especially for processes involving the $P$-wave Fock states. 
To solve the problem, the ``improved trace technology" is suggested and developed \cite{wbc1,cjx,lxz,Yang:2011ps}, which is based on the helicity amplitudes method and deals with the trace calculation directly at the amplitude level. 
In this way, the amplitudes could be expressed with the linear combinations of independent Lorentz structures.
In this paper, we adopt this technology to derive the analytical expression for all processes.

In previous works \cite{gxz1,gxz2}, the production of ground states ($1S$ and $1P$-wave) charmonium in $e^+e^-\to \gamma^*/Z^0 \to|(c\bar{c})\rangle +\gamma$ at super $Z$ factory is studied at the leading order and next-to-leading order in strong coupling constant $\alpha_s$ within NRQCD framework. 
The production of the ground states of both charmonium and bottomonium in $e^+e^-\to \gamma^*/Z^0 \to|(Q\bar{Q})\rangle +\gamma~(Q=c,b)$ at $Z^0$ peak are explored in Ref. \cite{Chang:2010am}, where the contribution from initial state radiation is also considered.
The production of the ground states of charmonium via virtual photon propagator in $e^+e^-\to \gamma^* \to|(c\bar{c})\rangle +\gamma$ at B-factories are discussed in system in Refs. \cite{Chung:2008km,Li:2009ki,Sang:2009jc}.
In the present paper, we shall concentrate our attention on the production of both ground and high Fock states of both charmonium and bottomonium in $e^+e^-\to \gamma^*/Z^0 \to|(Q\bar{Q})[n]\rangle +\gamma~(Q=c,b)$ at future super $Z$ factory, where $[n]$ is short for the color-singlet $[n^1S_0]$, $[n^3S_1]$, $[n^1P_0]$, and $[n^3P_J]$ Fock states ($n=1,2,3,4; J=0,1,2$). 
The analysis on differential distributions and the uncertainties shall be discussed.
This would be a helpful support for the experimental exploration on production of those high excited heavy charmonium and bottomnium at future super Z factory or GigaZ mode at CEPC.

In the literatures \cite{Liao:2012rh,lx,Liao:2015vqa,lx2}, we study the production of high excited heavy quarkonium in the decay of $W^\pm$, top quark, $Z^0$ and Higgs boson. 
The numerical results show that we can obtain sizable events of heavy quarkonium of high excited $[nS]$ and $[nP]$-wave states ($n \ge 2$), which implies that one can explore the special properties of those high excited states in experiments and should consider their contributions to the ground states properly. 
According to our study, in the processes of $e^+e^-\to \gamma^*/Z^0 \to|(Q\bar{Q})[n]\rangle +\gamma$, high excited sates could also be generated massively in comparison with the ground states.

The rest of the manuscript is organized as follows. 
In Section II, we introduce the calculation formalism and ``new trace technology" for the processes of $e^+e^-\to \gamma^*/Z^0 \to|(Q\bar{Q})[n]\rangle +\gamma$ within the NRQCD factorization framework. 
In Section III,  we evaluate the cross sections. The differential distributions of the cross sections and the uncertainties from various sources are studied in Sections \ref{production} and \ref{uncertainty}, respectively.
The final Section IV is reserved for a summary.

\section{Formulations and Calculation Techniques}
The cross sections for production of the charmonium in $e^+e^-\to \gamma^*/Z^0 \to |(c\bar{c})[n]\rangle +\gamma$ and bottomonium in $e^+e^-\to \gamma^*/Z^0 \to|(b\bar{b})[n]\rangle +\gamma$ can be calculated analogously under NRQCD factorization framework \cite{nrqcd1,nrqcd2}. 
The differential cross sections can be factored into the short-distance coefficients and the long-distance matrix elements,
\begin{equation} \label{dsigma}
d\sigma=\sum_{n} d\hat\sigma(|(Q\bar{Q})[n]\rangle) {\langle{\cal O}^H(n) \rangle}.
\end{equation} 
Here $\hat\sigma(|(Q\bar{Q})[n]\rangle)$ describes the short-distance production of a $(Q\bar{Q})$ pair ($Q=c$ or $b$ quarks) in the color, spin and angular momentum state $[n]$, and the non-perturbative NRQCD matrix elements $\langle{\cal O}^{H}(n)\rangle$ describe the hadronization of a Fock state $(Q\bar{Q})[n]$ into the heavy quarkonia $|(Q\bar{Q})[n]\rangle$. 
Here $[n]$ is short for $[n^1S_0]$, $[n^3S_1]$, $[n^1P_0]$ and $[n^3P_J]$ states with $n=1,2,3,4$ and $J=0,1,2$.
\begin{figure}
\includegraphics[width=0.4\textwidth]{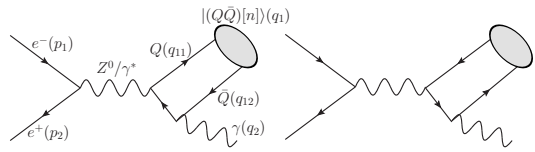}
\caption{Feynman diagrams for processes of $e^-(p_1) e^+(p_2) \to |(Q\bar{Q})[n]\rangle (q_1) + \gamma(q_2)$, where $|(Q\bar{Q})[n]\rangle$ stands for $|(c\bar{c})[n]\rangle$ and $|(b\bar{b})[n]\rangle$ quarkonia. 
Here $[n]$ is short for $[n^1S_0]$, $[n^3S_1]$, $[n^1P_0]$ and $[n^3P_J]$ Fock states with $n=1,2,3,4$ and $J=0,1,2$.} \label{feyn1}
\end{figure}

The short-distance differential cross section $d\hat\sigma$ are perturbatively calculable, and the two Feynman diagrams of the processes of $e^-(p_1)e^+(p_2)\to \gamma^*/Z^0 \to|(Q\bar{Q})[n]\rangle(q_1) +\gamma(q_2)$ are displayed in Fig.~\ref{feyn1}. 
Since the Feynman diagrams with initial state radiation can be identified in experiments, they are not considered here.
The perturbative differential cross section can be expressed as
\begin{eqnarray} \label{sd-sigma}
d\hat\sigma(|(Q\bar{Q})[n]\rangle)=\frac{1}{4\sqrt{(p_1\cdot p_2)^2-m^4_{e}}} \overline{\sum}  |{\cal M}(n)|^{2} d\Phi_2,
\end{eqnarray}
where $\overline{\sum}$ stands for the average over the spin of the initial particles and sum over the color and spin of the final particles when manipulating the squared amplitudes $|M(n)|^{2}$. 
In the $e^-e^+$ center-of-momentum (CM) frame, the two-body phase space can be simplified as
\begin{eqnarray}
    d{\Phi_2} &=& (2\pi)^4 \delta^{4}\left(p_1+p_2 - \sum_{f=1}^2 q_{f}\right)\prod_{f=1}^2 \frac{d^3{\vec{q}_f}}{(2\pi)^3 2q_f^0} \nonumber \\
    &=& \frac{{\mid\vec{q}_1}\mid}{8\pi\sqrt{s}}d(cos\theta).
    \label{dPhi-2}
\end{eqnarray}
In the second equation, we have made the integration over the $\delta$ function and the azimuth angle, and $\theta$ is the angle between the momentum $\vec{p_1}$ of electron and the momentum $\vec{q_1}$ of heavy quarkonium.
The parameter $s=(p_1+p_2)^2$ stands for the squared CM energy. 
The magnitude of the 3-dimension quarkonium momentum is ${|\vec{q}_1|}=(s-M^2_{Q\bar{Q}})/(2\sqrt{s})$, where $M_{Q\bar{Q}}$ is the mass of heavy quarkonium.

The hard scattering amplitude ${\cal M}(n)$ in Eq. (\ref{sd-sigma}) can be read directly from the Feynman diagrams in Fig.~\ref{feyn1}. 
And the general form of their amplitudes can be formulated as
\begin{eqnarray} \label{amplitude}
i{\cal M}(n)=  \sum_{k=1}^2 \bar{v}_{s'} (p_2) {\cal L}^\mu u_s (p_1)  {\cal D}_{\mu\nu} {\cal A}^\nu_k,
\end{eqnarray}
where the index $k$ represents the number of Feynman diagrams, and $s$ and $s'$ are the spins of the initial particles. 
The vertice $\cal{L}^\mu$ and the propagator $\cal{D_{\mu\nu}}$ for the virtual photon and $Z^0$ propagated processes have different forms,
\begin{eqnarray}
    \cal{L}^\mu &= \left\{
    \begin{array}{l}
        -i e \gamma^\mu \\
        \frac{-i g} {4 cos\theta_W} \gamma^\mu (1-4 sin^2 \theta_W-\gamma^5),
    \end{array}
    \right. \\
    \cal{D_{\mu\nu}} &= \left\{
    \begin{array}{l}
        \frac{-i g_{\mu\nu}}{k^2} \\
        \frac{i}{k^2-m^2_Z+i m_Z \Gamma_Z}(-g_{\mu\nu}+ \frac{k_\mu k_\nu }{k^2}). 
    \end{array}
    \right.
\end{eqnarray}
The upper and lower expressions after the big left bracket are for virtual photon and $Z^0$ propagated processes, respectively.
In which, $e$ is the unit of the electric charge, $g$ is the weak interaction coupling constant, $\theta_W$ represents the Weinberg angle, and $m_Z$ and $\Gamma_Z$ are the mass and the total decay width of $Z^0$ boson, respectively. 

The explicit expressions of the Dirac $\gamma$ matrix chains ${\cal A}^\nu_k$ in Eq. (\ref{amplitude}) for the $S$-wave spin-singlet $n^1S_0$ and spin-triplet $n^3S_1$ states ($n=1,2,3,4$) can be formulated as
\begin{widetext}
\begin{eqnarray}
{\cal A}^{\nu (S=0, L=0)}_1 &=& i Tr \left[ \Pi^{0}_{q_1}(q) {\cal R}^{\nu} \frac{-(\slashed{q}_2+ \slashed{q}_{12})+m_Q} {(q_2 + q_{12})^2-m^2_Q} \slashed{\epsilon} (q_2)\right] , \nonumber\\
{\cal A}^{\nu (S=0, L=0)}_2 &=& i  Tr \left[ \Pi^{0}_{q_1}(q) \slashed{\epsilon} (q_2) \frac{(\slashed{q}_2+ \slashed{q}_{11})+m_Q} {(q_2 + q_{11})^2-m^2_Q} {\cal R}^{\nu}\right] ; \nonumber\\
{\cal A}^{\nu (S=1, L=0)}_1 &=& i \epsilon_\alpha(q_1)  Tr \left[ \Pi^{\alpha}_{q_1}(q) {\cal R}^{\nu} \frac{-(\slashed{q}_2+ \slashed{q}_{12})+m_Q} {(q_2 + q_{12})^2-m^2_Q} \slashed{\epsilon} (q_2)\right] , \nonumber\\
{\cal A}^{\nu (S=1, L=0)}_2 &=& i \epsilon_\alpha(q_1)  Tr \left[ \Pi^{\alpha}_{q_1}(q)\slashed{\epsilon} (q_2)  \frac{(\slashed{q}_2+ \slashed{q}_{11})+m_Q} {(q_2 + q_{11})^2-m^2_Q}  {\cal R}^{\nu}\right] . \label{HadAmp-S}
\end{eqnarray}
\end{widetext}
The first two amplitudes are for $S$-wave spin-singlet states and the last two are for $S$-wave spin-triplet states. $\epsilon_\alpha (q_1)$ is the polarization vector for the spin-triplet states.
$\Pi^{0}_{q_1}(q)$ and $\Pi^{\alpha}_{q_1}(q)$ are the projectors for spin-singlet states and spin-triplet states respectively, with $q$ being the relative momentum between the two constituent quarks of heavy quarkonium. The two projectors have the following form 
\begin{eqnarray}
\Pi^{0}_{q_1}(q)=\frac{-1}{\sqrt{8 m^3_{Q}}}(\slashed{q}_{12}- m_{Q}) \gamma_5 (\slashed{q}_{11} + m_Q)\otimes  \frac{\delta_{ij}} {\sqrt{N_c}}, \nonumber\\
\Pi^{\alpha}_{q_1}(q)=\frac{-1}{\sqrt{8 m^3_{Q}}}(\slashed{q}_{12}- m_{Q}) \gamma_\alpha (\slashed{q}_{11} + m_Q)\otimes\frac{\delta_{ij}} {\sqrt{N_c}},
\end{eqnarray}
where $q_{11}=\frac{q_1}{2}+q$ and $q_{12}=\frac{q_1}{2}-q$ are the momenta of the two constituent heavy quarks, and $\delta_{ij}/\sqrt{N_c}$ is the color operator for color-singlet projector with $N_c=3$.
For the $S$-wave states, the relative momentum $q$ is set to zero directly.
The vertex $\cal{R}^\nu$ in Eq. (\ref{HadAmp-S}) is
\begin{equation}
    \cal{R}^\nu = \left\{
    \begin{array}{l}
        -i e e_{Q} \gamma^{\nu} \\
        \frac{-i g} {4 cos\theta_W} \gamma^\nu (1-4 e_Q  sin^2 \theta_W-\gamma^5)
    \end{array}
    \right.
\end{equation}
where the upper and lower expressions after the big left bracket are for the virtual photon and $Z^0$ propagated processes, respectively.
Here $e_Q=2/3$ for $c$ quark and $e_Q=-1/3$ for $b$ quark.

We now turn to the Dirac $\gamma$ matrix chains ${\cal A}^\nu_k$ in Eq. (\ref{amplitude}) for the $P$-wave spin-singlet $n ^1P_1$ and spin-triplet $n ^3P_J$ states ($n=1,2,3,4$), which can be expressed in terms of the $S$-wave ones in Eq. (\ref{HadAmp-S}),
\begin{eqnarray}
{\cal A}^{\nu (S=0, L=1)}_1 &=&  \epsilon_\beta(q_1) \left. \frac{d}{d q_\beta} {\cal A}^{\nu (S=0, L=0)}_1 \right|_{q=0}, \nonumber\\
{\cal A}^{\nu (S=0, L=1)}_2 &=&  \epsilon_\beta(q_1) \left. \frac{d}{d q_\beta} {\cal A}^{\nu (S=0, L=0)}_2 \right|_{q=0}; \nonumber\\
{\cal A}^{\nu (S=1, L=1)}_1 &=&  \varepsilon^J_{\alpha \beta}(q_1) \left. \frac{d}{d q_\beta} {\cal A}^{\nu (S=1, L=0)}_1 \right|_{q=0}, \nonumber\\
{\cal A}^{\nu (S=1, L=1)}_2 &=&  \varepsilon^J_{\alpha \beta}(q_1)  \left. \frac{d}{d q_\beta} {\cal A}^{\nu (S=1, L=0)}_2 \right|_{q=0}. \label{HadAmp-P}
\end{eqnarray}
The first two amplitudes are for $P$-wave spin-singlet states and the last two are for $P$-wave spin-triplet states.
In which, $\epsilon_\beta(q_1)$ is the polarization vector of the $n ^1P_1$ states and $\varepsilon^{J}_{\alpha\beta}(q_1)$ is the polarization tensor for $n^3P_J$ states with $J=0,1,2$. The derivatives over the relative momentum $q_\beta$ in Eq. (\ref{HadAmp-P}) will give complex and lengthy amplitudes.

When manipulating the squared amplitudes $|M(n)|^{2}$, we need to sum over the polarization vectors of the heavy quarknium. 
For the spin-triplet $n^3S_1$ states or the spin-singlet $n^1P_1$ states, the polarization sum is given by \cite{nrqcd2}
\begin{eqnarray}\label{3pja}
\sum_{J_z}\epsilon_{\alpha} \epsilon_{\alpha'} = \Pi_{\alpha\alpha'} \equiv -g_{\alpha \alpha'}+\frac{q_{1\alpha}q_{1\alpha'}}{M_{Q\bar{Q}}^{2}},
\end{eqnarray}
where $J_z=s_z$ or $l_z$ for $n^3S_1$ and $n^1P_1$ states, respectively. In the case of $n^3P_J$ states, the polarization sum should be performed by the selection of appropriate total angular momentum quantum number. 
The sum over polarization tensors  is given by \cite{nrqcd2}
\begin{eqnarray}\label{3pja}
\varepsilon^{(0)}_{\alpha\beta} \varepsilon^{(0)*}_{\alpha'\beta'} &=& \frac{1}{3} \Pi_{\alpha\beta}\Pi_{\alpha'\beta'}, \nonumber\\
\sum_{J_z}\varepsilon^{(1)}_{\alpha\beta} \varepsilon^{(1)*}_{\alpha'\beta'} &=& \frac{1}{2}
(\Pi_{\alpha\alpha'}\Pi_{\beta\beta'}- \Pi_{\alpha\beta'}\Pi_{\alpha'\beta}) \label{3pjb},\nonumber\\
\sum_{J_z}\varepsilon^{(2)}_{\alpha\beta} \varepsilon^{(2)*}_{\alpha'\beta'} &=& \frac{1}{2}
(\Pi_{\alpha\alpha'}\Pi_{\beta\beta'}+ \Pi_{\alpha\beta'}\Pi_{\alpha'\beta})-\frac{1}{3} \Pi_{\alpha\beta}\Pi_{\alpha'\beta'}, \nonumber\\ \label{3pjc}
\end{eqnarray}
for total angular momentum $J=0,1,2$, respectively.

To get compact analytical expression of the complicated $nP$-wave channels and also improve the efficiency of numerical evaluation, we adopt the ``improved trace technology" to simplify the amplitudes ${\cal M}(n)$ at the amplitude level before evaluating the polarization sum.
To shorten this manuscript, we present its main idea below.
For detailed techniques and more examples, one can refer to literatures \cite{wbc1,cjx,lxz,Yang:2011ps}.

Firstly, we introduce a massless spinor with negative helicity $u_-(k_0)$, which satisfies the following projection
\begin{equation}
{u_-(k_0)}{\bar{u}_-(k_0)}=\omega_{-} \slashed{k}_{0},
\end{equation}
where $k_0$ is an arbitrary light-like momentum, $k_0^2=0$, and $\omega_{-}=(1-\gamma_5)/{2}$. Then we construct the massless spinor with positive helicity
\begin{equation}
{u_+(k_0)}= \slashed{k}_{1}{u_-(k_0)},
\end{equation}
where $k_1$ is an arbitrary space-like momentum, $k_1^2=-1$, and satisfies $k_0\cdot k_1 =0$. It is easy to find that $u_+(k_0)$ has the projection relation
\begin{equation}
{u_+(k_0)}{\bar{u}_+(k_0)}=\omega_{+} \slashed{k}_{0},
\end{equation}
where $\omega_{+}=(1+\gamma_5)/{2}$.
Using these two massless spinors, one can construct the massive spionrs for the fermion and antifermion,
\begin{eqnarray}
{u_{\pm s}(p)} &=& (\slashed{p}+m){u_{\mp}(k_0)}/\sqrt{2{k_0}\cdot{p}},\nonumber\\
{v_{\pm s}(p)} &=& (\slashed{p}-m){u_{\mp}(k_0)}/\sqrt{2{k_0}\cdot{p}}.
\end{eqnarray}

Secondly, by using the above identities, one can write down the amplitude $M_{\pm{s}\pm{s'}}$ with four possible spin projections in the trace form directly 
\begin{eqnarray}
M_{ss'}&=&{N}Tr[(\slashed{p}_1+m_e) \omega_{-} \slashed{k}_0  (\slashed{p}_2-m_e) {A}],\nonumber\\
M_{-s-s'}&=&{N}Tr[(\slashed{p}_1+m_e) \omega_{+} \slashed{k}_0  (\slashed{p}_2-m_{e}) {A}],\nonumber\\
M_{-ss'}&=&{N}Tr[(\slashed{p}_1+m_e) \omega_{-} \slashed{k}_0\slashed{k}_1 (\slashed{p}_2-m_{e}) {A}],\nonumber\\
M_{s-s'}&=&{N}Tr[(\slashed{p}_1+m_e) \omega_{+} \slashed{k}_1 \slashed{k}_0 (\slashed{p}_2-m_{e}) {A}],\nonumber\\
\end{eqnarray}
where $A=\sum\limits_{k = 1}^{2} {\cal L}^\mu  {\cal D}_{\mu\nu} {\cal A}^\nu_k$ and the normalization constant $N = 1/\sqrt{4({k_0}\cdot{p_1})({k_0}\cdot{p_2})}$.
It is easy to check that $M_{\pm{s}\pm{s'}}$ are orthogonal for each other. Thus, the squared amplitude can be written as
\begin{eqnarray}
|M|^2 &=& |M_{ss'}|^2 + |M_{-s-s'}|^2 + |M_{-ss'}|^2 + |M_{s-s'}|^2\nonumber\\
&=& |M_{1}|^2 + |M_{2}|^2 + |M_{3}|^2 + |M_{4}|^2,
\end{eqnarray}
where we introduce four new amplitudes $M_i$ with ($i=1,\cdots,4$)
\begin{eqnarray}
M_1 &=& \frac{M_{ss'}+M_{-s-s'}}{\sqrt{2}}, M_2 = \frac{M_{ss'}-M_{-s-s'}}{\sqrt{2}}, \nonumber\\
M_3 &=& \frac{M_{s-s'}-M_{-ss'}}{\sqrt{2}}, M_4 = \frac{M_{s-s'}+M_{-ss'}}{\sqrt{2}}.
\end{eqnarray}

Thirdly, to obtain the explicit and compact expressions as much as possible, we choose $k_0 = {p_2} - \alpha {p_1}$ with $ \alpha = \left({p_2} \cdot {p_1} + \sqrt{({p_2} \cdot {p_1})^2 - {p^2_2} {p^2_1}}\right)/{p^2_1}$, and $k_1^\mu  = i{N_0}{\varepsilon^{\mu \nu \rho \sigma }}{p_{1\nu}}{q_{1\rho}}{p_{2\sigma}}$, which leads to
\begin{eqnarray}
\slashed{k}_1 = {N_0}{\gamma _5} \left[ ({p_1} \cdot {q_1}) {\slashed{p}_2} + ({q_1} \cdot {p_2}) {\slashed{p}_1} - ({p_1} \cdot {p_2}) \slashed{q}_1 
 -\slashed{p}_1 \slashed{q}_1 \slashed{p}_2 \right]. \nonumber
\end{eqnarray}
Then the amplitudes $M_i$ can be expressed as
\begin{eqnarray} \label{trace-M}
M_1 &=& {L_1} \times Tr [(\slashed{p}_1+m_e)  (\slashed{p}_2-m_{e})  A],\nonumber\\
M_2 &=& {L_2} \times Tr [(\slashed{p}_1+m_e) {\gamma _5}  (\slashed{p}_2-m_{e})  A],\nonumber\\
M_3 &=& M_{3'}-{N_0}[({p_2}\cdot{q_2}){m_e}+({p_1} \cdot {q_2}) m_{e}]M_2, \nonumber\\
M_4 &=& M_{4'}+{N_0}[({p_2}\cdot{q_2}){m_e}-({p_1} \cdot {q_2}) m_{e}]M_1,
\end{eqnarray}
where $L_{1,2}=1/(2\sqrt{p_1\cdot p_2 \mp m_e^2})$ and
\begin{eqnarray}
M_{3'} &=&\frac{N_0}{4L_2} Tr\left[ {(\slashed{p}_1+m_e)  {\gamma _5}  {\slashed{q}_2}  ((\slashed{p}_2-m_{e}) A } \right],\nonumber\\
M_{4'} &=-&\frac{N_0}{4L_1} Tr\left[ {(\slashed{p}_1+m_e)  {\slashed{q}_2}  (\slashed{p}_2-m_{e})  A } \right] .
\end{eqnarray}
The normalization factor $N_0$ is determined by ensuresing $k_1\cdot k_1=-1$.
Thus after the three steps above, the amplitudes $M_i$ in Eq. (\ref{trace-M}) would be expressed by the linear combinations of some independent Lorentz structures. 

We finally discuss the non-perturbative matrix elements ${\langle{\cal O}^H(n) \rangle}$ in Eq. (\ref{dsigma}). They can be calculated through the lattice QCD \cite{lat1}, the potential NRQCD \cite{pnrqcd1,yellow}, or the potential models \cite{lx,Eichten:1978tg,Eichten:1979ms,pot2,pot3,pot4,Chen:1992fq,Eichten:1994gt}. 
In this manuscript, we adopt the potential models to describe the non-perturbative hadronization of a $(Q\bar{Q})[n]$ Fock state into the heavy quarkonium $|(Q\bar{Q})[n]\rangle$.
For color-singlet Fock states, the matrix elements are related to the Schr\"{o}dinger wave function at the origin $\Psi_{\mid(Q\bar{Q})[nS]\rangle}(0)$ for the $nS$-wave Fock states, or the first derivative of the wave function at the origin $\Psi^\prime_{\mid(Q\bar{Q})[nP]\rangle}(0)$ for the $nP$-wave states \cite{nrqcd1},
\begin{eqnarray}
\langle{\cal O}^H(nS) \rangle &\simeq& |\Psi_{\mid(Q\bar{Q})[nS]\rangle}(0)|^2,\nonumber\\
\langle{\cal O}^H(nP) \rangle &\simeq& |\Psi^\prime_{\mid(Q\bar{Q})[nP]\rangle}(0)|^2.
\end{eqnarray}
Due to the fact that the spin-splitting effects are small, the same values of wave function for both the spin-singlet and spin-triplet Fock states are adopted in our calculation. 
Further, the Schr\"{o}dinger wave function at the origin $\Psi_{|Q\bar{Q})[nS]\rangle}(0)$ and its first derivative at the origin $\Psi^{'}_{|(Q\bar{Q})[nP]\rangle}(0)$ are related to the radial wave function at the origin $R_{|(Q\bar{Q})[nS]\rangle}(0)$ and its first derivative at the origin $R^{'}_{|(Q\bar{Q})[nP]\rangle}(0)$, respectively \cite{nrqcd1},
\begin{eqnarray}
\Psi_{|(Q\bar{Q})[nS]\rangle}(0)&=&\sqrt{{1}/{4\pi}}R_{|(Q\bar{Q})[nS]\rangle}(0),\nonumber\\
\Psi'_{|(Q\bar{Q})[nP]\rangle}(0)&=&\sqrt{{3}/{4\pi}}R'_{|(Q\bar{Q})[nP]\rangle}(0).
\end{eqnarray}
Note that if one would take the color-octet Fock states into consideration, the color-octet NRQCD matrices are suppressed by certain orders in $v_Q$ to the corresponding color-singlet ones based on the velocity scale rules of NRQCD \cite{nrqcd1,Brambilla:2010cs,Wu:2002ig}.
One can also derive the values of color-octet NRQCD matrix elements by fitting the experimental measurements \cite{Ma:2010yw,Ma:2010jj}.

\section{Numerical Results}
\subsection{Input parameters}
\begin{table}
\caption{Masses (units: GeV) of the constituent quark and radial wave functions at the origin $|R_{|(Q\bar{Q})[nS]\rangle}(0)|^2$ (units: GeV$^3$) and their first derivatives at the origin $|R'_{|(Q\bar{Q})[nP]\rangle}(0)|^2$ (units: GeV$^5$) within the BT-potential model~\cite{lx}. 
Uncertainties of radial wave functions at the origin and their first derivatives at the origin 
are caused by the corresponding varying quark masses.}
\begin{tabular}{|c|c|c|}
\hline
~&~$m_c$,~~~$|R_{|(c\bar{c})[nS]\rangle}(0)|^2$~&~$m_c$,~~~$|R'_{|(c\bar{c})[nP]\rangle}(0)|^2$\\
\hline
$n=1$ &~1.48$\pm$0.1~,~$2.458^{+0.227}_{-0.327}$~&~1.75$\pm$0.1~,~$0.322^{+0.077}_{-0.068}$~\\
$n=2$ &~1.82$\pm$0.1~,~$1.671^{+0.115}_{-0.107}$~&~1.96$\pm$0.1~,~$0.224^{+0.012}_{-0.012}$~\\
$n=3$ &~1.92$\pm$0.1~,~$0.969^{+0.063}_{-0.057}$~&~2.12$\pm$0.1~,~$0.387^{+0.045}_{-0.042}$~\\
$n=4$ &~2.02$\pm$0.1~,~$0.796^{+0.064}_{-0.054}$~&~2.26$\pm$0.1~,~$0.467^{+0.057}_{-0.053}$~\\
\hline\hline
~&~$m_b$,~~~$|R_{|(b\bar{b})[nS]\rangle}(0)|^2$~&~$m_b$,~~~$|R'_{|(b\bar{b})[nP]\rangle}(0)|^2$\\
\hline
$n=1$ &~4.71$\pm$0.2~,~$16.12^{+1.28}_{-1.23}$~&~4.94$\pm$0.2~,~$5.874^{+0.728}_{-0.675}$~\\
$n=2$ &~5.01$\pm$0.2~,~$6.746^{+0.598}_{-0.580}$~&~5.12$\pm$0.2~,~$2.827^{+0.492}_{-0.432}$~\\
$n=3$ &~5.17$\pm$0.2~,~$2.172^{+0.178}_{-0.155}$~&~5.20$\pm$0.2~,~$2.578^{+0.187}_{-0.186}$~\\
$n=4$ &~5.27$\pm$0.2~,~$2.588^{+0.110}_{-0.114}$~&~5.37$\pm$0.2~,~$3.217^{+0.283}_{-0.271}$~\\
\hline\hline
\end{tabular}
\label{M&R}
\end{table}

In our numerical analysis, the quark mass $m_Q$ is set to be half the mass of heavy quarkonium $M_{Q\bar{Q}}/2$, which ensures the gauge invariance of the hard scattering amplitude under the NRQCD framework. 
The masses of $c$ and $b$ quarks for the ground and high excited quarkonia are displayed in Table \ref{M&R}.
In our previous work \cite{lx}, we calculate the radial wave functions at the origin $R_{|(Q\bar{Q})[nS]\rangle}(0)$ and the first derivatives of radial wave functions at the origin $R'_{|(Q\bar{Q})[nP]\rangle}(0)$ for heavy quarkonium $|(c\bar{c})[n]\rangle,~|(b\bar{c})[n]\rangle$ and $|(b\bar{b})[n]\rangle$ under five different potential models.  
In this work, we use the results of the Buchm\"{u}ller and Tye potential model (BT-potential) \cite{pot2,wgs}, which are also presented in Table \ref{M&R}.
We will discuss the uncertainties from the radial wave functions at the origin and their derivatives at the origin under different potential models in Section \ref{uncertainty}.
Note that in Table \ref{M&R}, the uncertainties of radial wave functions at the origin and their first derivatives at the origin are caused by the corresponding varying quark masses.
It tells us that the evaluation of cross sections of high excited Fock states ($n=2,3,4$) are more than simply replacing the non-perturbative matrix elements in the calculation for the ground state ($n=1$).
The non-perturbative matrix elements depend on the heavy quark masses.
Other parameters have the following values \cite{pdg}: the mass of $Z^0$ boson $m_Z =91.1876$ GeV and its total decay width $\Gamma_{Z^0}=2.4952$ GeV, the Fermi constant $G_F=\frac{\sqrt{2}g^2}{8m_W^2}=1.16639 \times 10^{-5}$ GeV$^{-2}$ with $m_W= 80.399$ GeV, the Weinberg angle $\theta_W=\arcsin\sqrt{0.23119}$, and the fine structure constant $\alpha=e^2/4\pi=1/130.9$.

\subsection{Heavy quarkonium production in $e^-e^+\to \gamma^*/Z^0\to|(Q\bar{Q})[n]\rangle+\gamma$}
\label{production}
\begin{table}
\caption{Cross sections (units: $\times 10^{-4}fb$) for $e^-e^+\to\gamma^*\to|(Q\bar{Q})[n]\rangle+\gamma$ at $\sqrt{s}=91.1876$ GeV under the BT-potential model. Percentages in brackets are ratios relative to the ground state.}
\begin{tabular}{|c|c|c|c|c|c|}
\hline
$\gamma^*,~|(Q\bar{Q})[n]\rangle$&~$n=1$~&~$n=2$~&~$n=3$~&~$n=4$~\\
\hline\hline
$\sigma{(|(c\bar{c})[n^1S_0]\rangle)}$&~413.6~&~221.1(53\%)~&~125.6(30\%)~&~98.05(24\%)~\\
\hline
$\sigma{(|(c\bar{c})[n^3P_0]\rangle)}$&~14.89~&~7.365(49\%)~&~10.01(67\%)~&~9.963(67\%)~\\
\hline
$\sigma{(|(c\bar{c})[n^3P_1]\rangle)}$&~90.28~&~44.77(49\%)~&~61.00(68\%)~&~60.82(67\%)~\\
\hline
$\sigma{(|(c\bar{c})[n^3P_2]\rangle)}$&~30.18~&~14.98(50\%)~&~20.43(68\%)~&~20.37(68\%)~\\
\hline
Sum&~549.0~&~288.3(53\%)~&~217.0(40\%)~&~189.2(34\%)~\\
\hline\hline
$\sigma{(|(b\bar{b})[n^1S_0]\rangle)}$&~52.76~&~20.73(39\%)~&~6.462(12\%)~&~7.550(14\%)~\\
\hline
$\sigma{(|(b\bar{b})[n^3P_0]\rangle)}$&~0.715~&~0.308(43\%)~&~0.267(37\%)~&~0.302(42\%)~\\
\hline
$\sigma{(|(b\bar{b})[n^3P_1]\rangle)}$&~4.664~&~2.019(43\%)~&~1.759(38\%)~&~1.997(43\%)~\\
\hline
$\sigma{(|(b\bar{b})[n^3P_2]\rangle)}$&~1.592~&~0.691(43\%)~&~0.602(38\%)~&~0.685(43\%)~\\
\hline
Sum&~59.73~&~23.75(40\%)~&~9.091(15\%)~&~10.53(18\%)~\\
\hline\hline
\end{tabular}
\label{aXS}
\end{table}
\begin{table}
\caption{Ccross sections (units: $\times 10^{-2}fb$) for $e^-e^+\to Z^0 \to |(Q\bar{Q})[n]\rangle+\gamma$ at $\sqrt{s}=91.1876$ GeV under the BT-potential model. Percentages in brackets are ratios relative to the ground state.}
\begin{tabular}{|c|c|c|c|c|c|}
\hline
$Z^0,~|(Q\bar{Q})[n]\rangle$&~$n=1$~&~$n=2$~&~$n=3$~&~$n=4$~\\
\hline\hline
$\sigma{(|(c\bar{c})[n^1S_0]\rangle)}$&~239.8~&~128.2(53\%)~&~72.82(30\%)~&~56.86(24\%)~\\
\hline
$\sigma{(|(c\bar{c})[n^3S_1]\rangle)}$&~1632~&~873.2(53\%)~&~496.0(30\%)~&~387.3(24\%)~\\
\hline
$\sigma{(|(c\bar{c})[n^1P_1]\rangle)}$&~177.1~&~87.71(50\%)~&~119.8(68\%)~&~119.3(67\%)~\\
\hline
$\sigma{(|(c\bar{c})[n^3P_0]\rangle)}$&~8.620~&~4.261(49\%)~&~5.807(67\%)~&~5.776(67\%)~\\
\hline
$\sigma{(|(c\bar{c})[n^3P_1]\rangle)}$&~52.26~&~25.89(50\%)~&~35.38(68\%)~&~35.26(67\%)~\\
\hline
$\sigma{(|(c\bar{c})[n^3P_2]\rangle)}$&~17.47~&~8.664(50\%)~&~11.84(68\%)~&~11.81(68\%)~\\
\hline
Sum&~2128~&~1128(53\%)~&~741.6(35\%)~&~616.3(29\%)~\\
\hline\hline
$\sigma{(|(b\bar{b})[n^1S_0]\rangle)}$&~398.1~&~156.4(39\%)~&~48.76(12\%)~&~56.96(14\%)~\\
\hline
$\sigma{(|(b\bar{b})[n^3S_1]\rangle)}$&~840.8~&~330.8(39\%)~&~103.2(12\%)~&~120.6(14\%)~\\
\hline
$\sigma{(|(b\bar{b})[n^1P_1]\rangle)}$&~35.91~&~15.52(43\%)~&~13.51(38\%)~&~{15.31(43\%)}~\\
\hline
$\sigma{(|(b\bar{b})[n^3P_0]\rangle)}$&~5.395~&~2.322(43\%)~&~2.017(37\%)~&~2.275(42\%)~\\
\hline
$\sigma{(|(b\bar{b})[n^3P_1]\rangle)}$&~35.19~&~15.24(43\%)~&~13.27(38\%)~&~15.07(43\%)~\\
\hline
$\sigma{(|(b\bar{b})[n^3P_2]\rangle)}$&~12.01~&~5.210(43\%)~&~4.543(38\%)~&~5.165(43\%)~\\
\hline
Sum&~1328~&~525.5(40\%)~&~185.3(14\%)~&~215.4(15\%)~\\
\hline\hline
\end{tabular}
\label{ZXS}
\end{table}

The total cross sections for the production of heavy quarkonia via $e^-e^+\to \gamma^*/Z^0\to|(Q\bar{Q})[n]\rangle+\gamma$ ($Q=c \text{~or~} b$ quarks) at center-of-momentum (CM) energy $\sqrt{s}=91.1876$ GeV are listed in Tables \ref{aXS} and \ref{ZXS} for virtual photon $\gamma^*$ and $Z^0$ propagated processes, respectively. 
The percentages in brackets are ratios of high excited states ($n=2,3,4$) relative to the ground state ($n=1$).
Here we adopt the BT-potential model to evaluate the non-perturbative hadronic matrix elements~\cite{lx}. 
It is worth noting that, there are no estimations on the $\sigma(|(Q\bar{Q})[n^3S_1]\rangle)$ and $\sigma(|(Q\bar{Q})[n^1P_1]\rangle)$ via the virtual photon propagated processes in Table \ref{aXS} because they break up the conservation of $C$ parity.
In Refs. \cite{gxz1,gxz2}, Chen {\it et. al.} calculate the cross sections for $1S$ and $1P$-wave charmonium in $e^-e^+\to \gamma^*/Z^0\to|(c\bar{c})[n]\rangle+\gamma$ at leading and next-to-leading order accuracy in strong coupling constant $\alpha_s$. If the same input parameters are adopted, our estimations are consistent with theirs at leading order.

Since the units in Table \ref{ZXS} are two orders larger than units in Table \ref{aXS}, the contributions from the virtual photon processes are negligible at future super Z factory. In Table \ref{ZXS} for $Z^0$ propagated processes, it is found that
\begin{eqnarray}
&&\sigma(|(Q\bar{Q}[n^3S_1]\rangle) >\sigma(|(Q\bar{Q}[n^1S_0]\rangle), \nonumber\\
&&\sigma(|(Q\bar{Q}[n^1P_1]\rangle) > \sigma(|(Q\bar{Q}[n^3P_1]\rangle)>\sigma(|(Q\bar{Q}[n^3P_2]\rangle)\nonumber\\ &&> \sigma(|(Q\bar{Q}[n^3P_0]\rangle),
\end{eqnarray}
where $Q=c$ or $b$ quarks.
For bottomonium $|b\bar{b}[n]\rangle$, the cross sections of $n^1P_1$ Fock state for all $n=1,2,3,4$ are quite close to those of the $n^3P_1$ Fock state at the same $n$th level. 
It is worth noting that in Ref. \cite{Chang:2010am}, they considered the contribution from initial state radiation and found that $\sigma(|(b\bar{b}[1^3P_2]\rangle) > \sigma(|(b\bar{b}[1^3P_1]\rangle) > \sigma(|(b\bar{b}[1^1P_1]\rangle) > \sigma(|(b\bar{b}[1^3P_0]\rangle)$ as shown in Table 2 therein. 
Their estimates for  $\sigma(|(b\bar{b}[1^1P_1]\rangle)$ and $\sigma(|(b\bar{b}[1^3P_1]\rangle)$ are also quite close.
The relations of magnitudes for charmonium are consistent with each other.

Let's take a closer look at the cross sections of the high excited states in Table \ref{ZXS}.
When using $[nS]$ to represent the sum of cross sections of $n^1S_0$ and $n^3S_1$, and $[nP]$ to represent the sum of cross sections of $n^1P_1$ and $n^3P_J$ ($J=0,1,2$) at the same $n$th level, we have
\begin{itemize}
\item For $|(c\bar{c})[n]\rangle$ quarkonium, the cross sections for $2S$, $3S$, $4S$, $1P$, $2P$, $3P$ and $4P$-wave states are about $53.5\%$, $30.4\%$, $23.7\%$, $13.7\%$, $6.8\%$, $9.2\%$, and $9.2\%$ of the cross section of the $|(c\bar{c})[1S]\rangle$ quarkonium, respectively.
\end{itemize}
\begin{itemize}
\item For $|(b\bar{b})[n]\rangle$ quarkonium, the cross sections for $2S$, $3S$, $4S$, $1P$, $2P$, $3P$ and $4P$-wave states are about $39.3\%$, $12.3\%$, $14.3\%$, $7.1\%$, $3.1\%$, $2.7\%$, and $3.1\%$ of the cross section of the $|(b\bar{b})[1S]\rangle$ quarkonium, respectively.
\end{itemize}
Then at the future $Z$ factory or CEPC in GigaZ mode running at CM energy $\sqrt{s}=m_Z$ with high luminosity, we can obtain sizable events to study both ground and high excited heavy quarkonia. We can obtain the events in one operation year simply by multiplying the cross sections in Tables \ref{aXS} and \ref{ZXS} by the luminosity ${\cal L}\approx 10^{36}cm^{-2}s^{-1} \approx 10^4 fb^{-1} year^{-1}$. 

\begin{figure*}[htbp]
\centering
\includegraphics[width=0.32\textwidth]{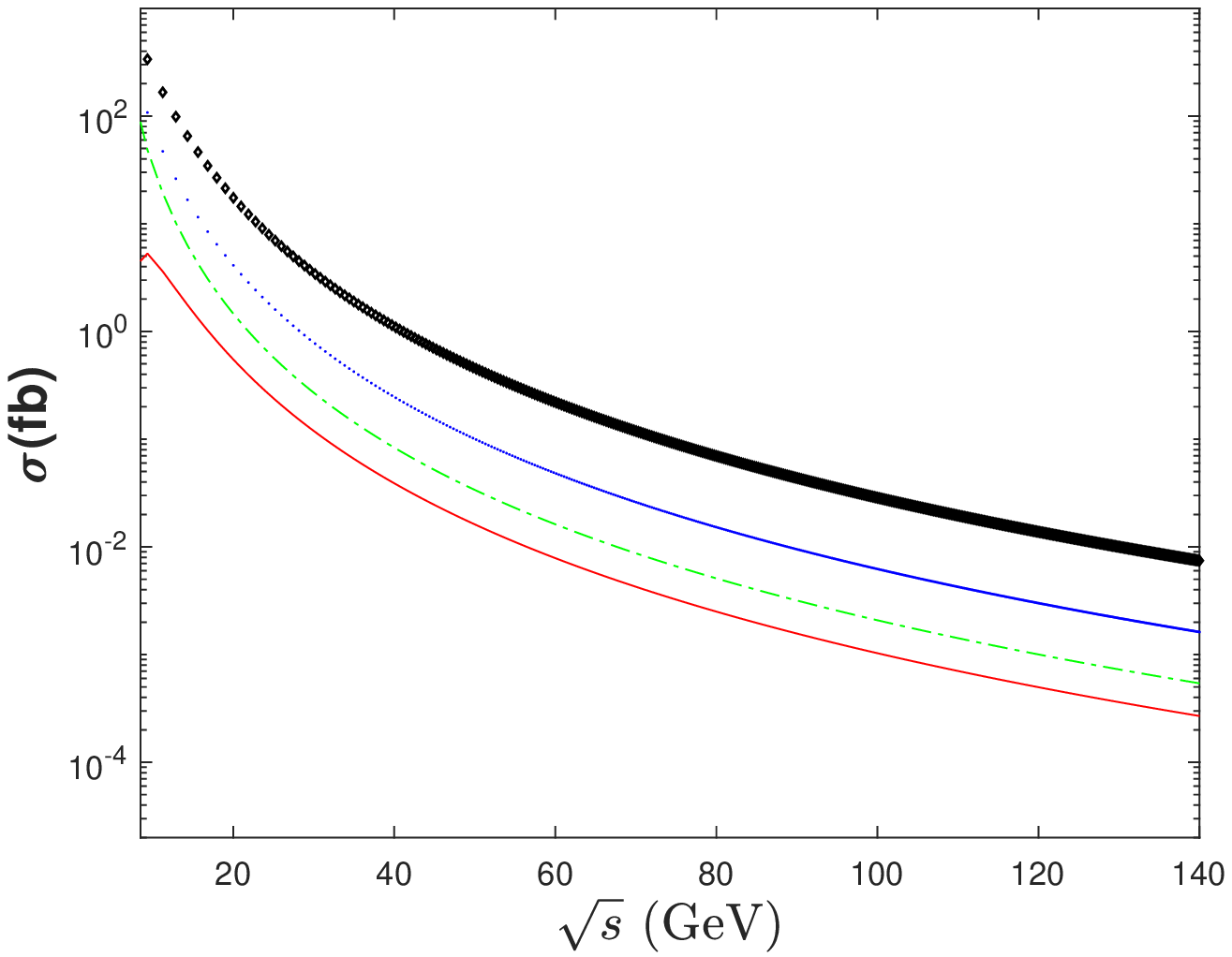}
\includegraphics[width=0.32\textwidth]{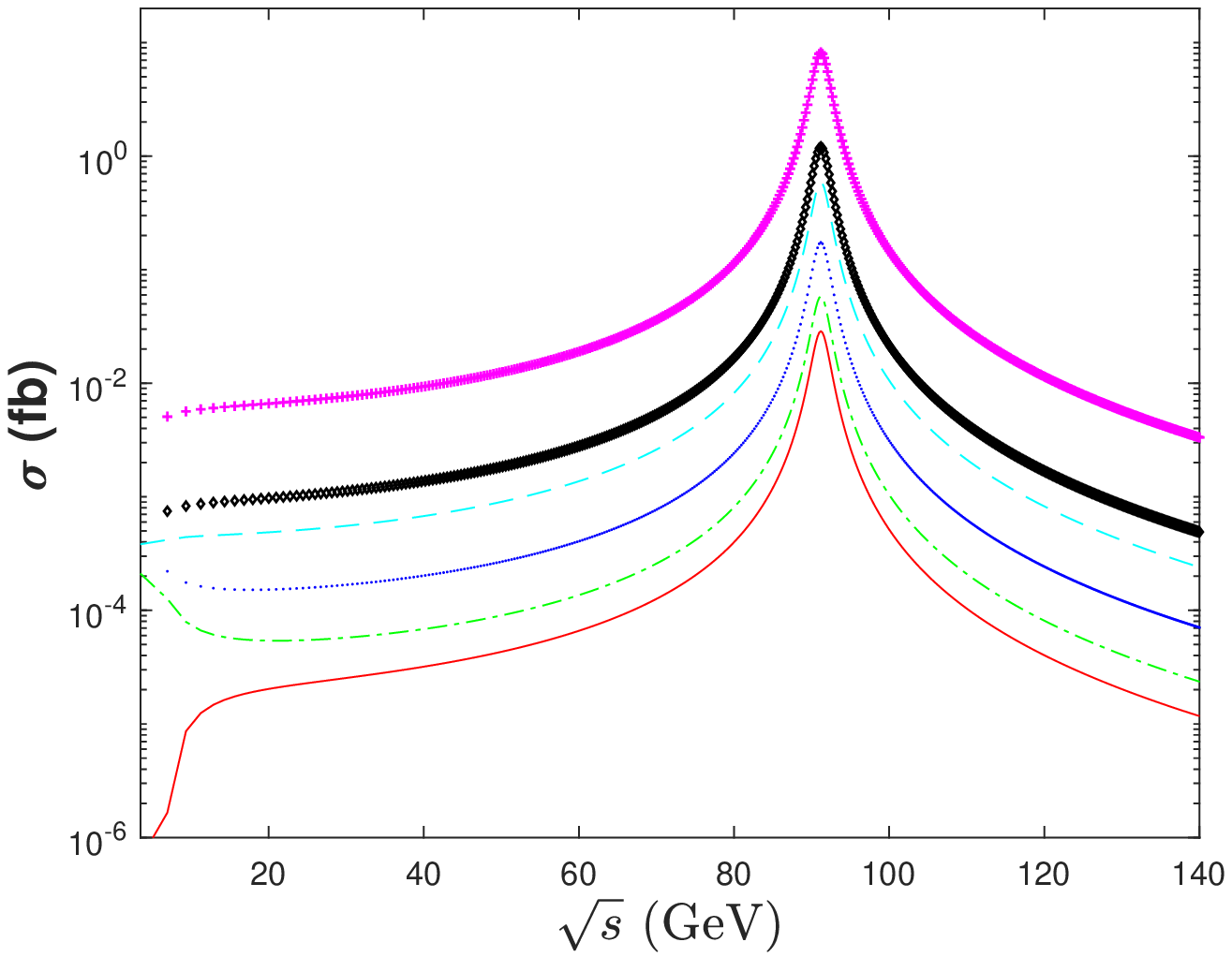}
\includegraphics[width=0.32\textwidth]{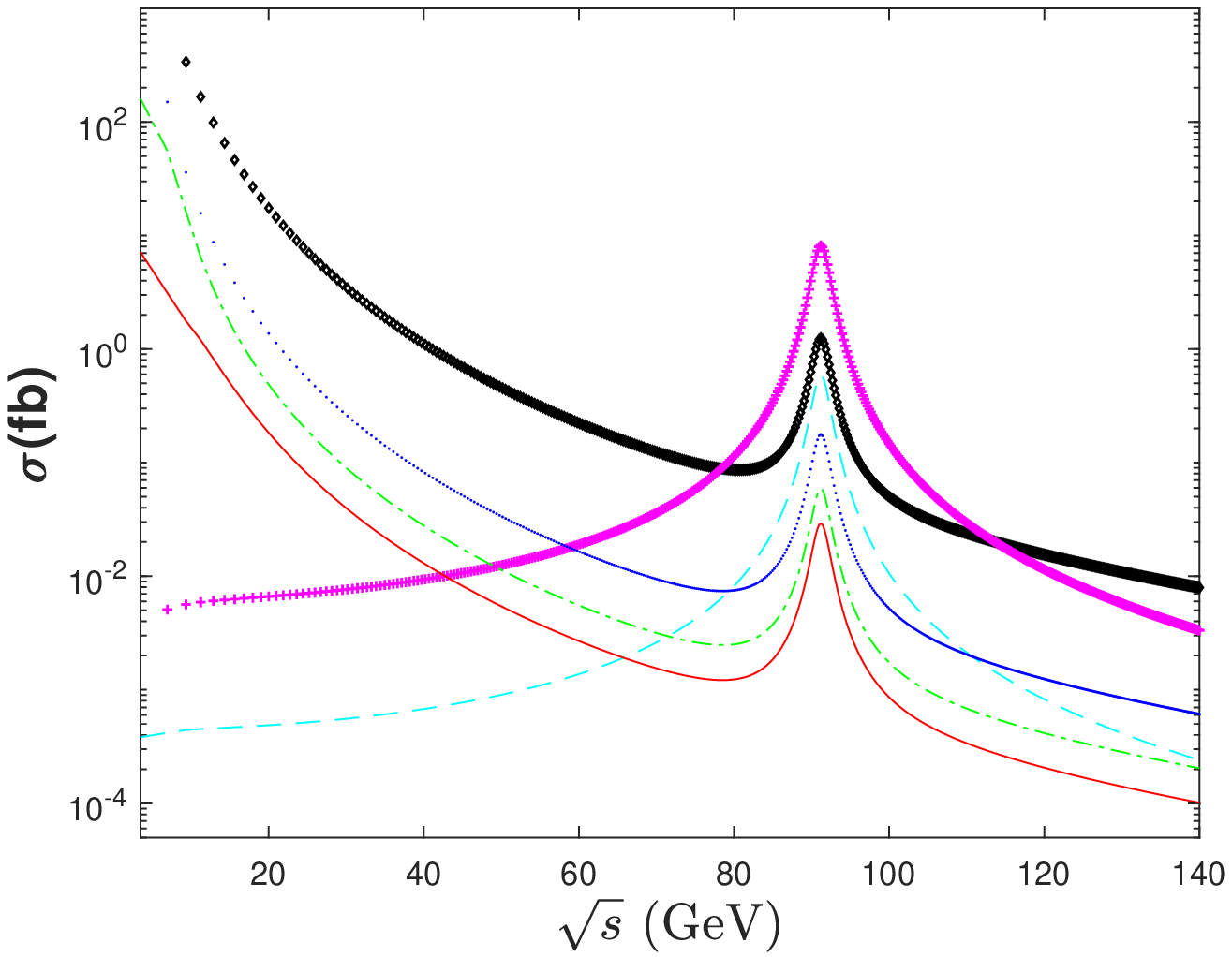}
\caption{
Cross sections versus the CM energy $\sqrt{s}$ for the channel $e^-e^+\to \gamma^*/Z^0 \to|(c\bar{c})[1]\rangle+\gamma$ via the virtual photon $\gamma^*$ (left), the $Z^0$ boson (middle) and the sum of previous two (right). The diamond black line, cross magenta line, dashed cyan line, solid red line, dotted blue line, and the dash-dotted green line are for $|(c\bar{c})[1^1S_0]\rangle$, $|(c\bar{c})[1^3S_1]\rangle$, $|(c\bar{c})[1^1P_1]\rangle$, $|(c\bar{c})[1^3P_0]\rangle$, $|(c\bar{c})[1^3P_1]\rangle$, $|(c\bar{c})[1^3P_2]\rangle$, respectively.
} \label{ccds}
\end{figure*}
\begin{figure*}[htbp]
\centering
\includegraphics[width=0.32\textwidth]{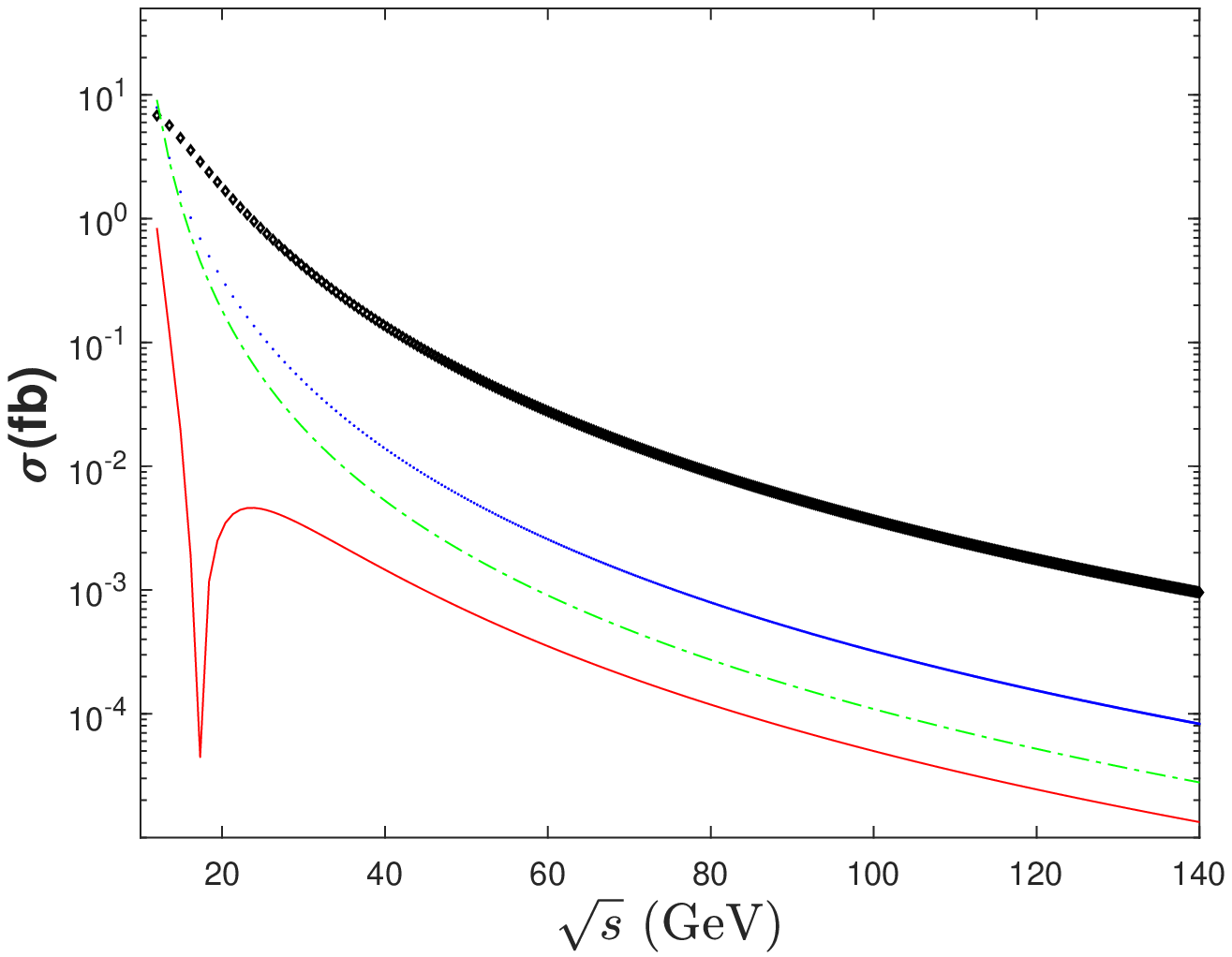}
\includegraphics[width=0.32\textwidth]{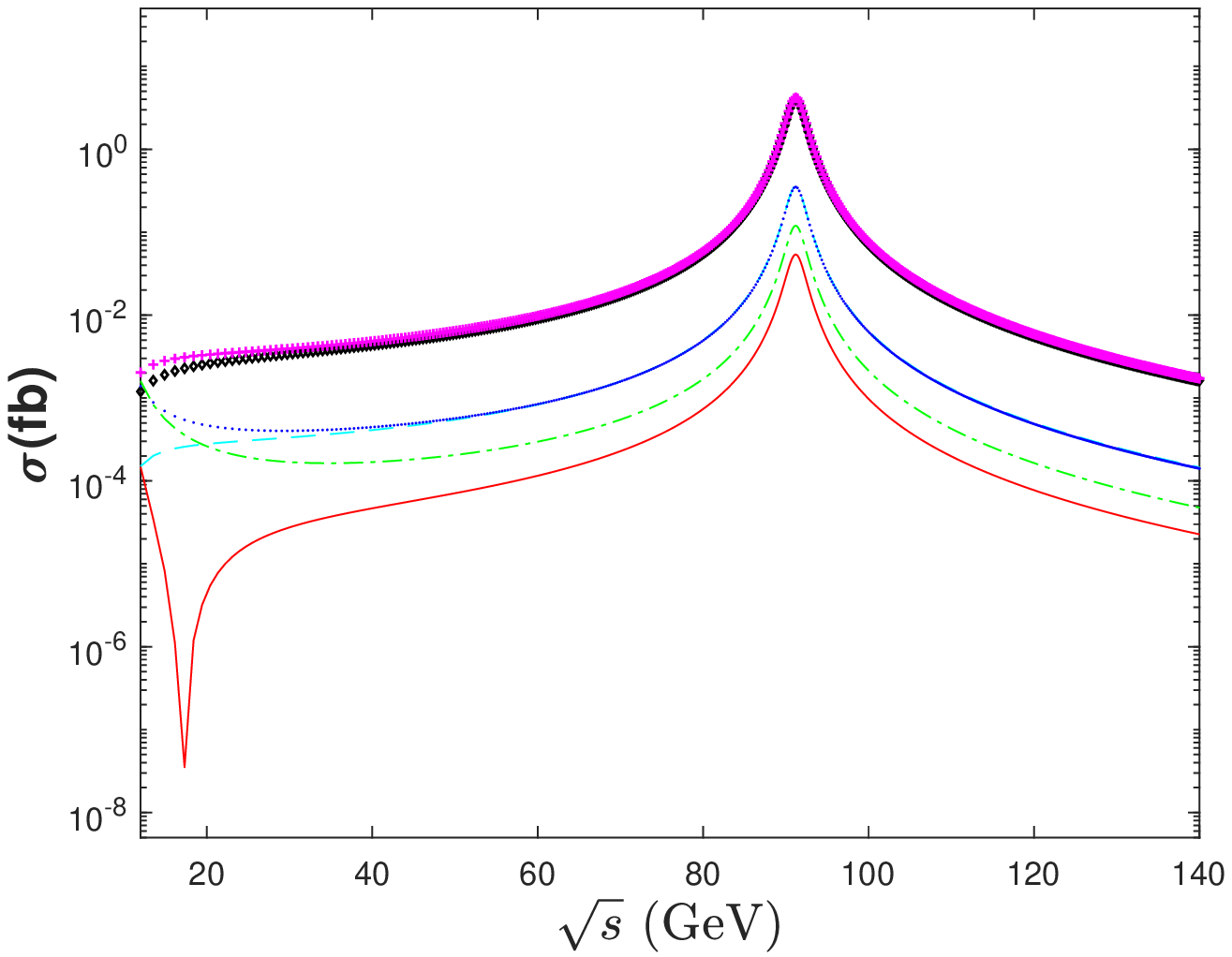}
\includegraphics[width=0.32\textwidth]{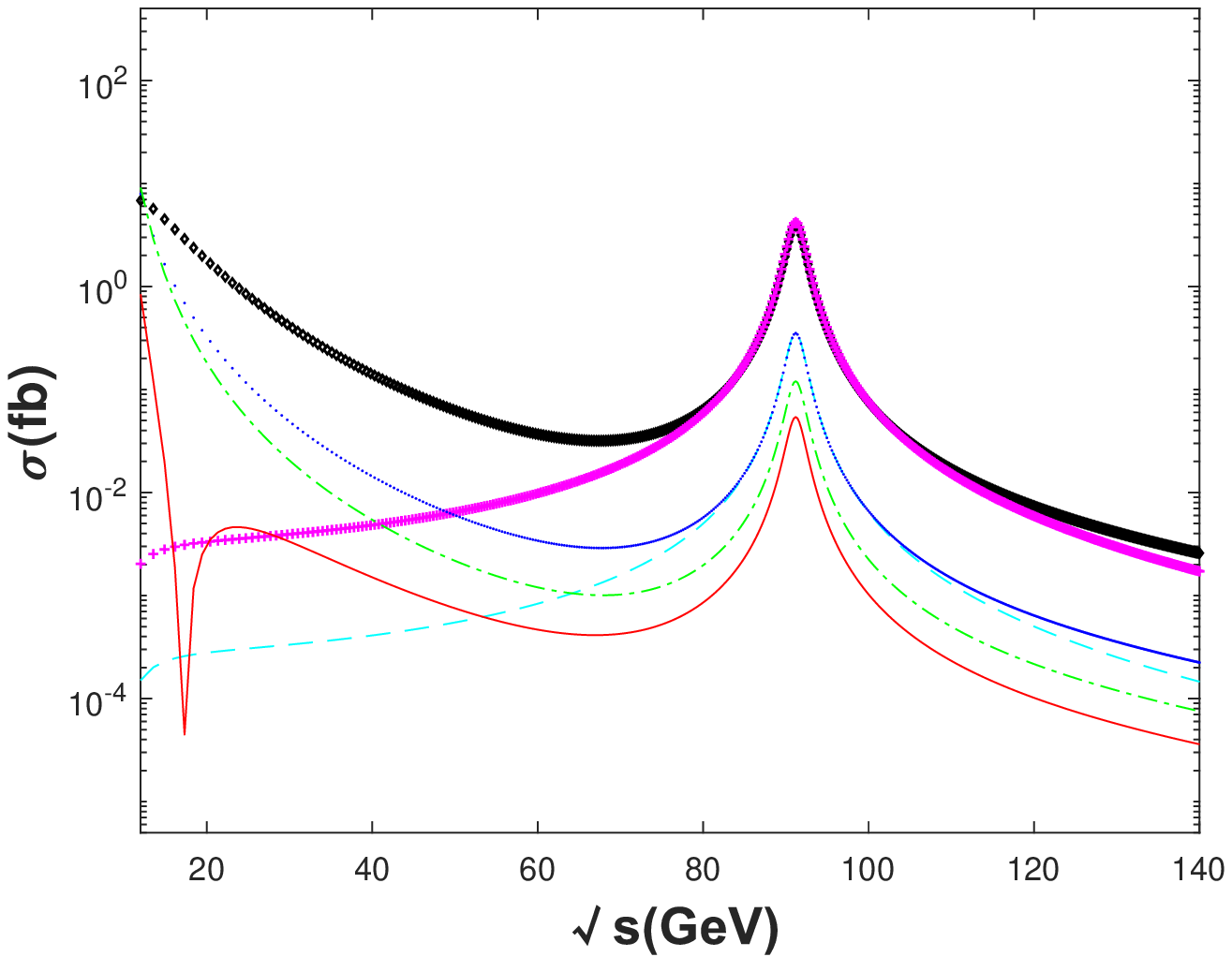}
\caption{
Cross sections versus the CM energy $\sqrt{s}$ for the channel $e^-e^+\to \gamma^*/Z^0 \to|(b\bar{b})[1]\rangle+\gamma$ via the virtual photon $\gamma^*$ (left), the $Z^0$ boson (middle) and the sum of previous two (right). The diamond black line, cross magenta line, dashed cyan line, solid red line, dotted blue line, and the dash-dotted green line are for $|(b\bar{b})[1^1S_0]\rangle$, $|(b\bar{b})[1^3S_1]\rangle$, $|(b\bar{b})[1^1P_1]\rangle$, $|(b\bar{b})[1^3P_0]\rangle$, $|(b\bar{b})[1^3P_1]\rangle$, $|(b\bar{b})[1^3P_2]\rangle$, respectively.
} \label{bbds}
\end{figure*}

In  Figs. \ref{ccds} and \ref{bbds}, we display the total cross sections versus the CM energy $\sqrt{s}$ for ground states $|(c\bar{c})[1]\rangle$ and $|(b\bar{b})[1]\rangle$ respectively, where $[1]$ stands for $1^1S_0$,  $1^3S_1$,  $1^1P_1$, and $1^3P_J$-wave states ($J=0,1,2$).
They show explicitly the contributions of $\gamma^*$ and $Z^0$ propagated processes from $\sqrt{s}=10$ GeV to 140 GeV.
Around the $Z^0$ peak, the $Z^0$ propagated processes dominate without any doubts.
The curves of total cross sections versus $\sqrt{s}$ for high excited states $|(c\bar{c})[n]\rangle$ and $|(b\bar{b})[n]\rangle$ with $n=2,3,4$ have similar line shapes.

\begin{figure*}[htbp]
\subfigure[$\gamma^*,~|(c\bar{c}){[1]}\rangle$
]{
\begin{minipage}[t]{0.25\linewidth}
\centering
\includegraphics[width=1.0\textwidth]{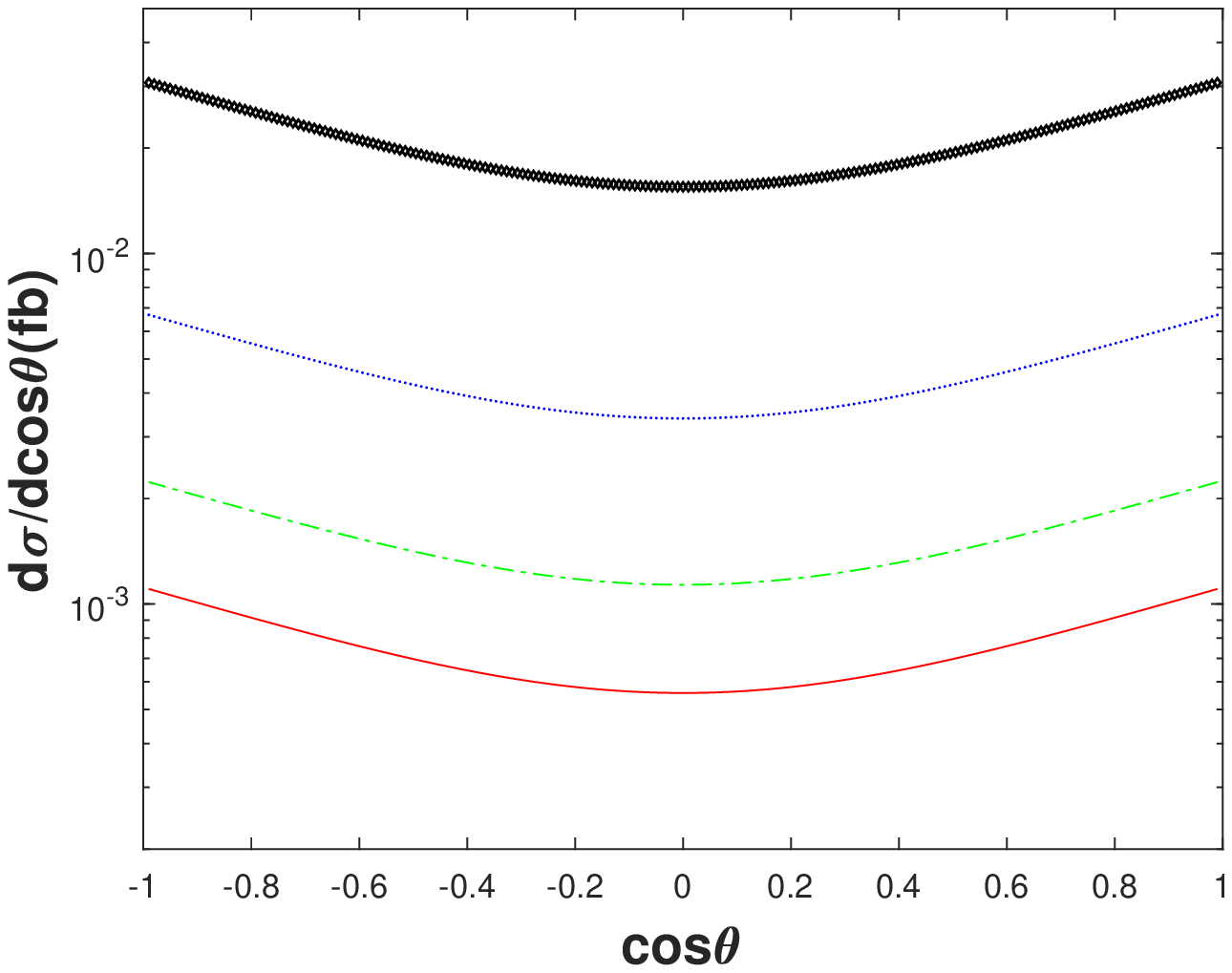}
\end{minipage}%
}%
\subfigure[$Z^0,~|(c\bar{c}){[1]}\rangle$
]{
\begin{minipage}[t]{0.25\linewidth}
\centering
\includegraphics[width=1.0\textwidth]{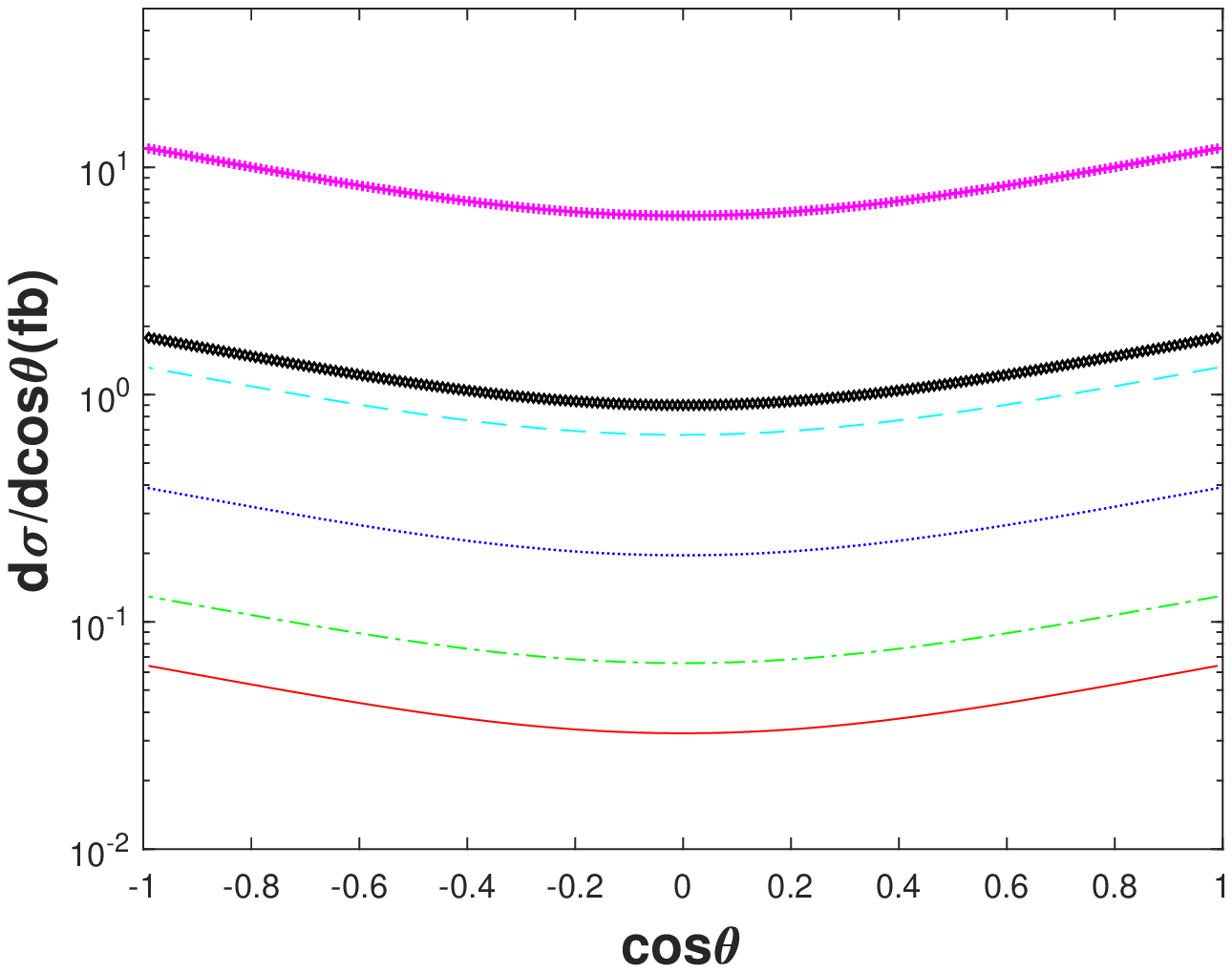}
\end{minipage}%
}%
\subfigure[$\gamma^*,~|(b\bar{b}){[1]}\rangle$
]{
\begin{minipage}[t]{0.25\linewidth}
\centering
\includegraphics[width=1.0\textwidth]{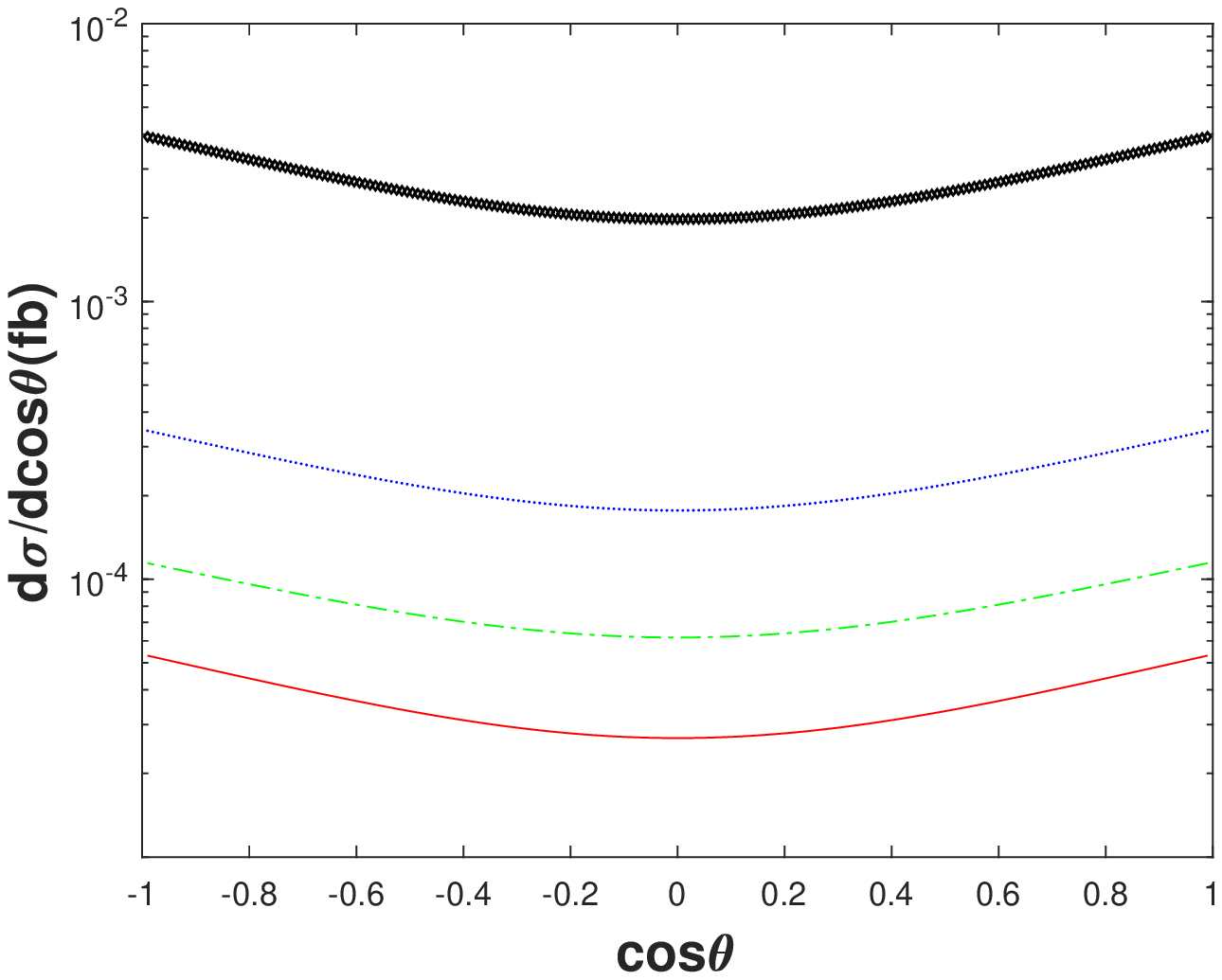}
\end{minipage}%
}%
\subfigure[$Z^0,~|(b\bar{b}){[1]}\rangle$
]{
\begin{minipage}[t]{0.25\linewidth}
\centering
\includegraphics[width=1.0\textwidth]{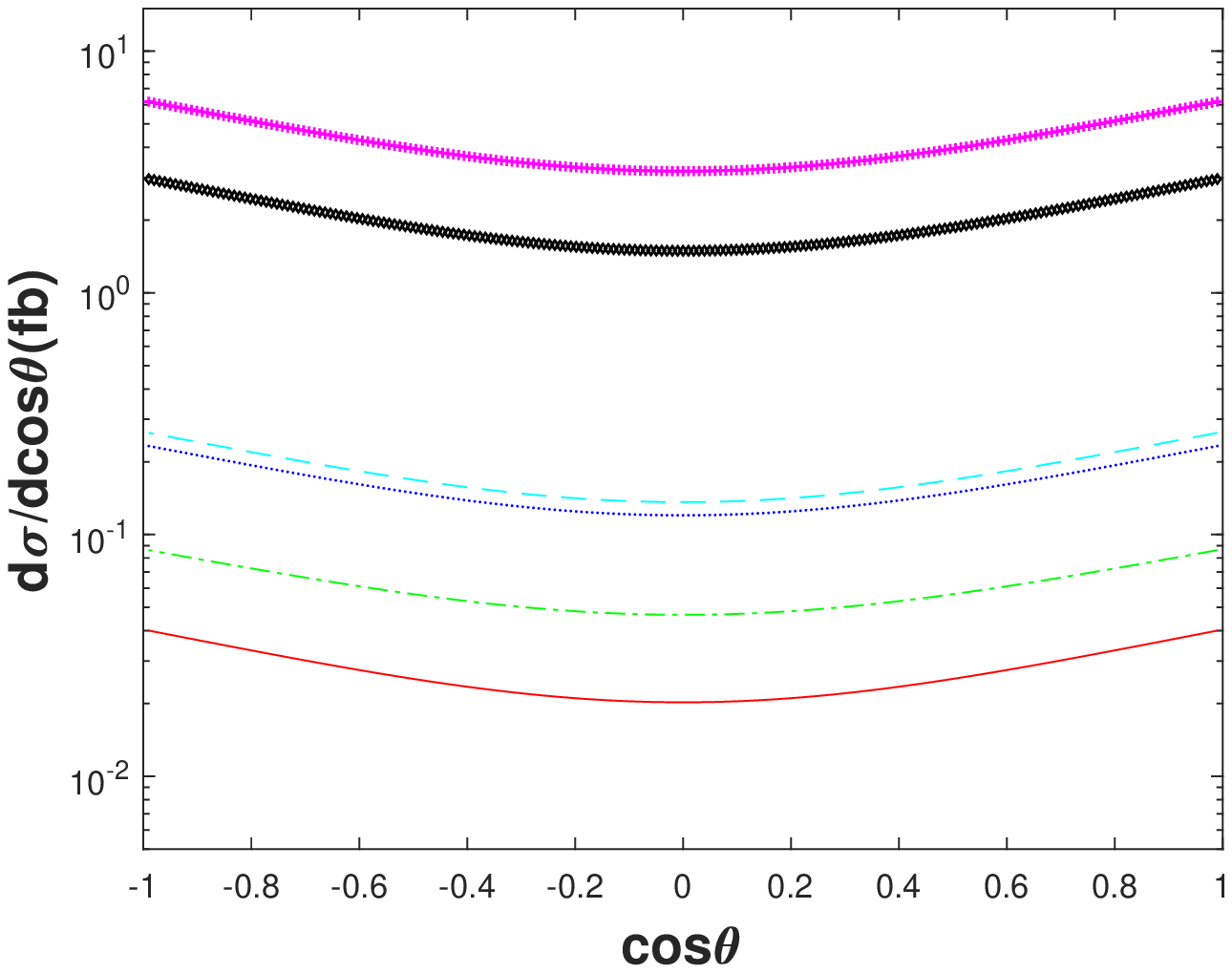}
\end{minipage}%
}%
\caption{
Differential cross sections d$\sigma$/dcos$\theta$ for: (a) $|(c\bar{c})[1]\rangle$ via $\gamma^*$ propagator, (b) $|(c\bar{c})[1]\rangle$ via $Z^{0}$ propagator, (c) $|(b\bar{b})[1]\rangle$ via $\gamma^*$ propagator, (d) $|(b\bar{b})[1]\rangle$ via $Z^{0}$ propagator. The diamond black line, cross magenta line, dashed cyan line, solid red line, dotted blue line, and the dash-dotted green line are for $1^1S_0$, $1^3S_1$, $1^1P_1$, $1^3P_0$, $1^3P_1$, $1^3P_2$, respectively.
} \label{ccbbdcos}
\end{figure*}

In Fig. \ref{ccbbdcos}, differential distributions $d\sigma/dcos\theta$ for ground states $|(c\bar{c})[1]\rangle$ and $|(b\bar{b})[1]\rangle$ are displayed, where $[1]$ stands for $1^1S_0$,  $1^3S_1$,  $1^1P_1$, and $1^3P_J$-wave states ($J=0,1,2$). 
Here, $\theta$ is the angle between the momentum $\vec{p_1}$ of electron and the momentum $\vec{q_1}$ of the heavy quarkonium.
It is shown that the $Z^0$ propagated processes and the corresponding virtual photon propagated ones have similar line shapes.
We also find that $d\sigma/dcos\theta$ approaches its maximum when the heavy quarkonium and the electron running in the same direction or back-to-back for both $S$-wave and $P$-wave states.
The curves of differential cross sections $d\sigma/dcos\theta$ for high excited states $|(c\bar{c})[n]\rangle$ and $|(b\bar{b})[n]\rangle$ with $n=2,3,4$ have similar line shapes.
 
\begin{figure*}[htbp]
\subfigure[$\gamma^*,~|(c\bar{c}){[1]}\rangle$
]{
\begin{minipage}[t]{0.25\linewidth}
\centering
\includegraphics[width=1.0\textwidth]{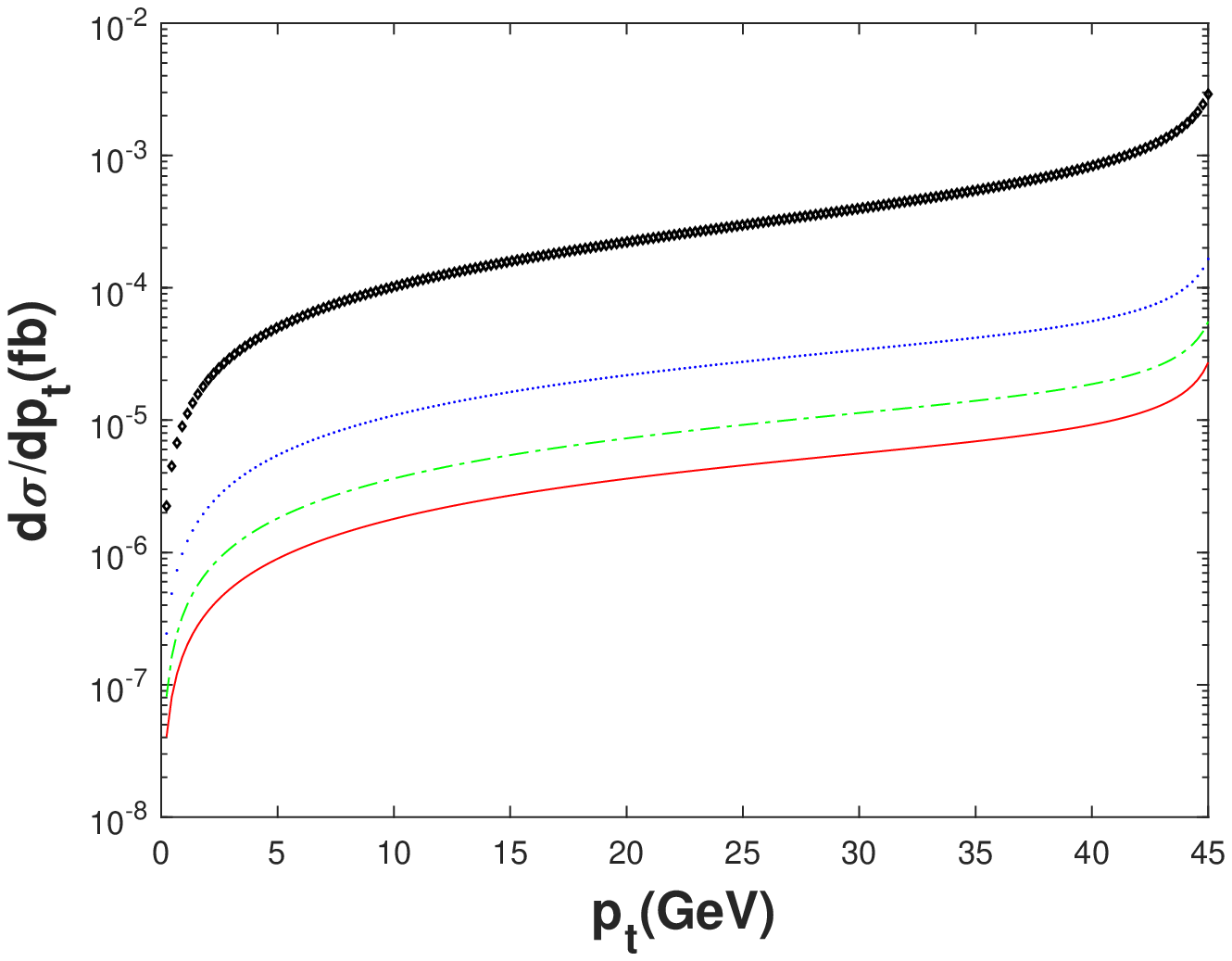}
\end{minipage}%
}%
\subfigure[$Z^0,~|(c\bar{c}){[1]}\rangle$
]{
\begin{minipage}[t]{0.25\linewidth}
\centering
\includegraphics[width=1.0\textwidth]{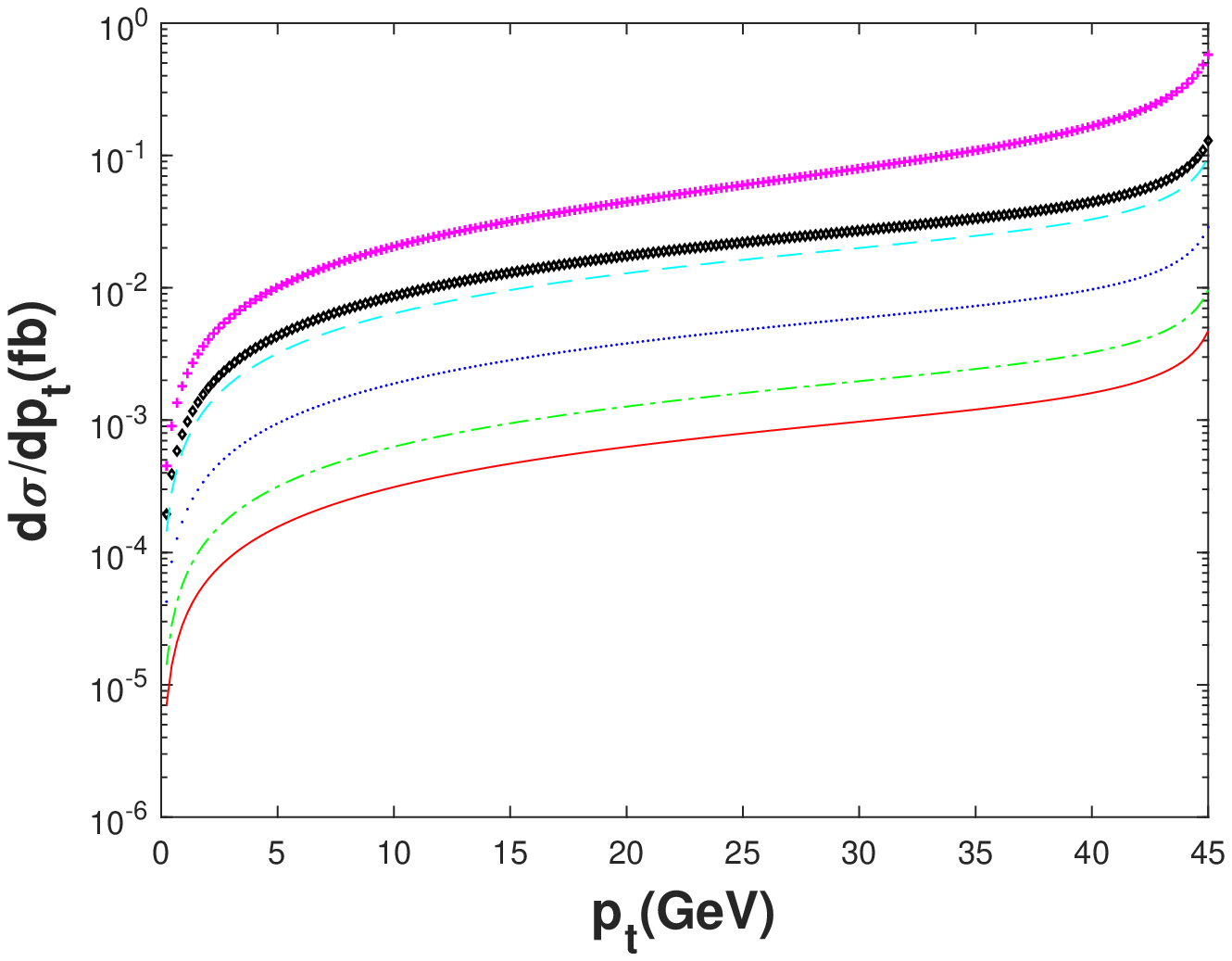}
\end{minipage}%
}%
\subfigure[$\gamma^*,~|(b\bar{b}){[1]}\rangle$
]{
\begin{minipage}[t]{0.25\linewidth}
\centering
\includegraphics[width=1.0\textwidth]{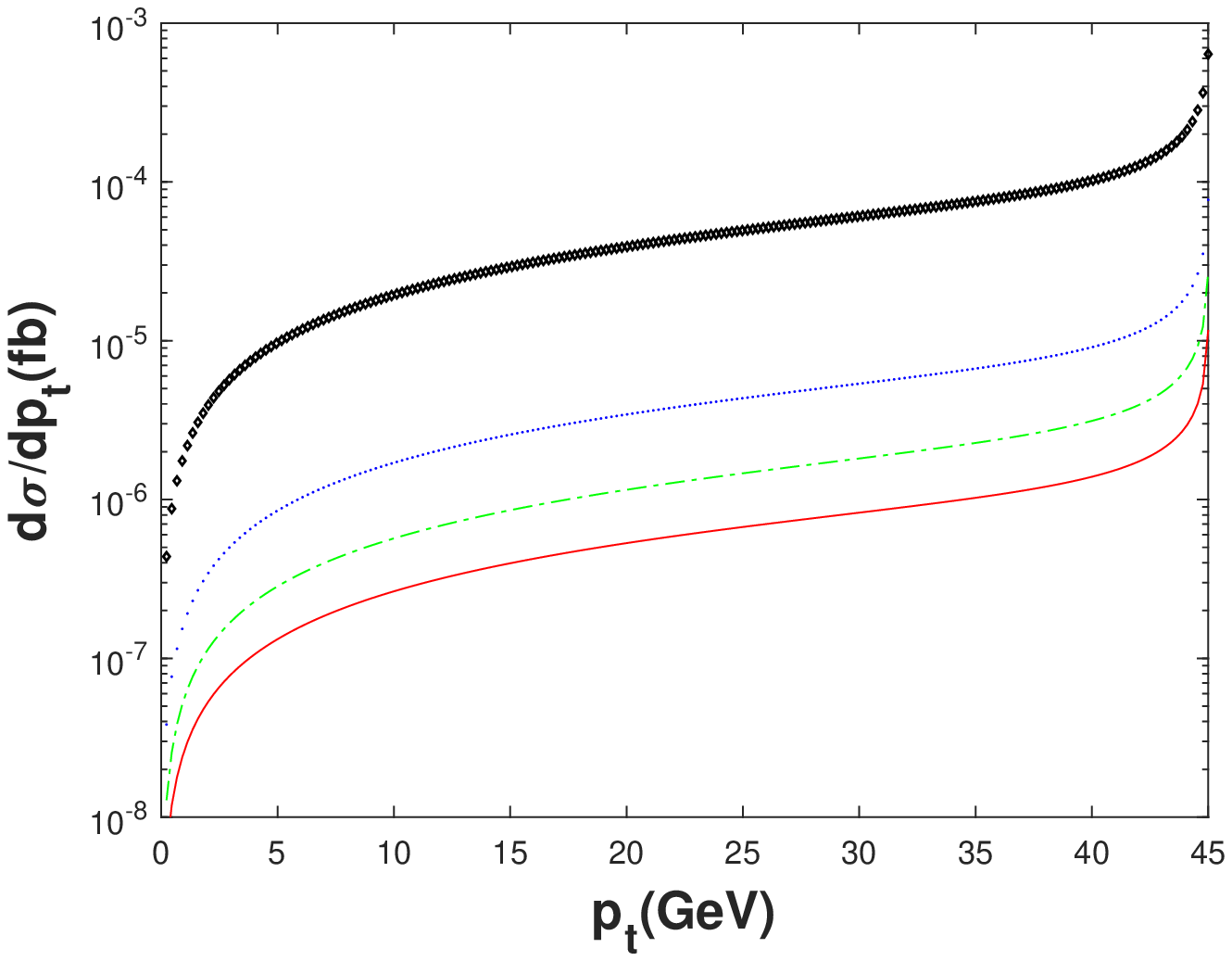}
\end{minipage}%
}%
\subfigure[$Z^0,~|(b\bar{b}){[1]}\rangle$
]{
\begin{minipage}[t]{0.25\linewidth}
\centering
\includegraphics[width=1.0\textwidth]{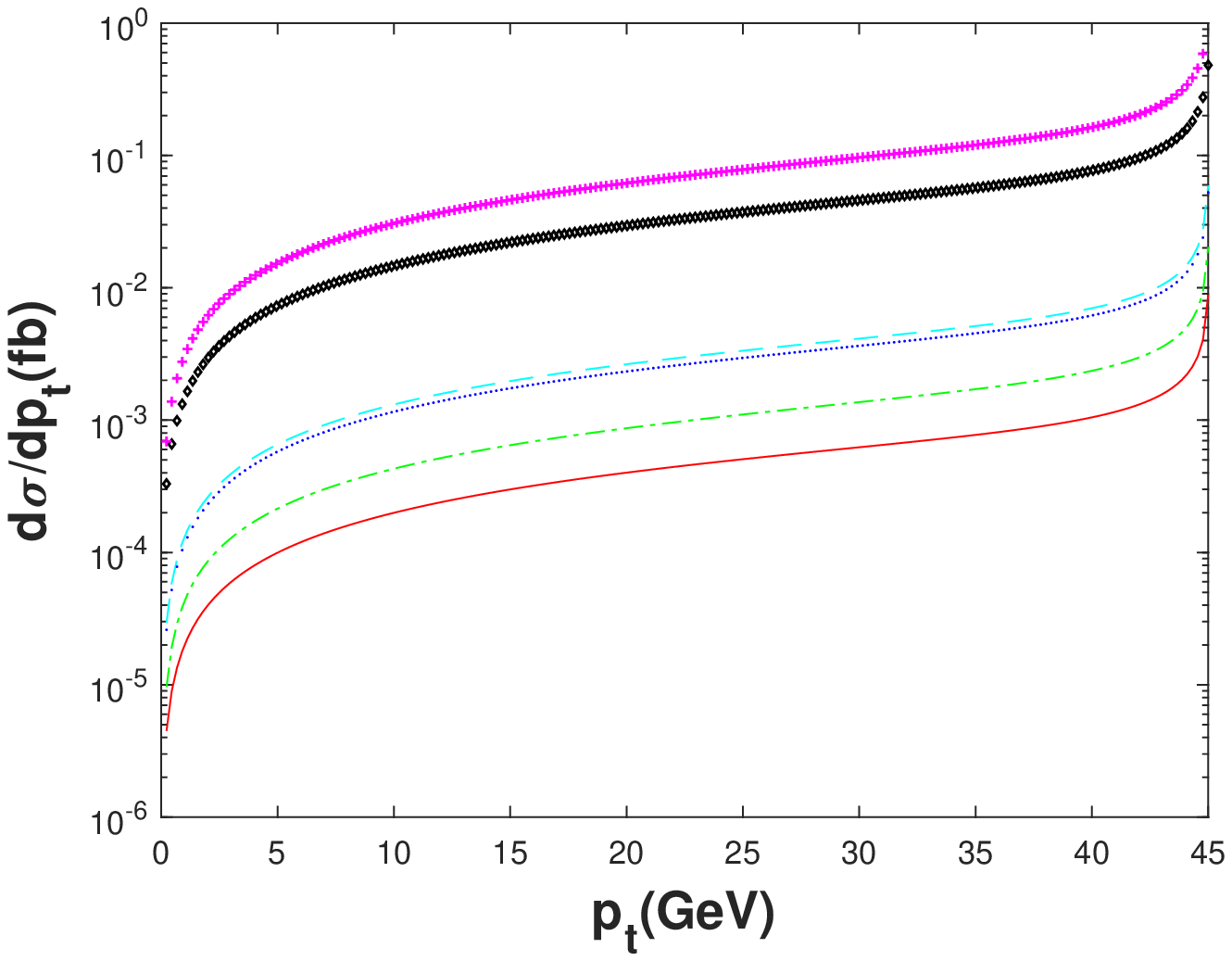}
\end{minipage}%
}%
\caption{
Differential cross sections d$\sigma$/d$p_t$ for: (a) $|(c\bar{c})[1]\rangle$ via $\gamma^*$ propagator, (b) $|(c\bar{c})[1]\rangle$ via $Z^0$ propagator, (c) $|(b\bar{b})[1]\rangle$ via $\gamma^*$ propagator, (d) $|(b\bar{b})[1]\rangle$ via $Z^0$ propagator. The diamond black line, cross magenta line, dashed cyan line, solid red line, dotted blue line, and the dash-dotted green line are for $1^1S_0$, $1^3S_1$, $1^1P_1$, $1^3P_0$, $1^3P_1$, $1^3P_2$, respectively.
} \label{ccbbdpt}
\end{figure*}
The transverse momentum $p_{t}$ distribution of the heavy quarkonium can further tell us more information on the production of the charmonium and bottomonium.
If the distribution $d\sigma/dcos\theta$ is set to be
\begin{eqnarray}
\frac{d\sigma}{dcos\theta}=f(cos\theta),
\end{eqnarray}
which can be easily obtained with the differential phase space of Eq. (\ref{dPhi-2}), then the distribution $d\sigma/dp_t$ can be obtained by
\begin{eqnarray}
\frac{d\sigma}{dp_t}&=&\left|\frac{dcos\theta}{dp_t}\right| \left(\frac{d\sigma}{dcos\theta}\right)\nonumber\\
&=&\frac{p_t}{|\vec{q}_1| \sqrt{|\vec{q}_1|^2-p^2_t}} f(cos\theta),
\end{eqnarray}
where $|\vec{q}_1|=(s-M^2_{Q\bar{Q}})/(2\sqrt{s})$ is the magnitude of the momentum of the heavy quarkonium. 
We present the transverse momentum $p_t$ distributions for the cross sections in Fig. \ref{ccbbdpt} for ground states $|(c\bar{c})[1]\rangle$ and $|(b\bar{b})[1]\rangle$. 
Since the differential distribution is proportional to $p_t/\sqrt{|\vec{q}_1|^2-p^2_t}$ and values of the function $f(cos\theta)$ changes smoothly, $d\sigma/dp_t$ shall increase with the increment of transverse momentum $p_t$.
The curves of differential cross sections $d\sigma/dp_t$ for high excited states $|(c\bar{c})[n]\rangle$ and $|(b\bar{b})[n]\rangle$ with $n=2,3,4$ have similar line shapes.

\subsection{Uncertainty analysis}
\label{uncertainty}
\begin{table}
\caption{Uncertainties of total cross sections (units: $\times 10^{-4}$ fb) caused by varying the masses as shown in Table \ref{M&R} for $\gamma^*$ propagated processes. Note, effects of uncertainties of radial wave functions at the origin and their first derivatives at the origin caused by varying masses are also considered.}
\begin{tabular}{|c|c|c|c|c|}
\hline
$|(Q\bar{Q})[n]\rangle$&~$n=1$~&~$n=2$~&~$n=3$~&~$n=4$~\\
\hline\hline
$\sigma_{(|(c\bar{c})[n^1S_0]\rangle)}$&$413.6^{+9.5}_{-29.0}$&$221.1^{+3.4}_{-2.6}$&$125.6^{+1.5}_{-0.7}$&$98.05^{+12.13}_{-11.00}$\\
$\sigma_{(|(c\bar{c})[n^3P_0]\rangle)}$&$14.89^{+0.72}_{-0.84}$&$7.365^{+0.804}_{-0.680}$&$10.01^{+0.31}_{-0.28}$&$9.963^{+0.162}_{-0.157}$\\
$\sigma_{(|(c\bar{c})[n^3P_1]\rangle)}$&$90.29^{+4.46}_{-5.18}$&$44.77^{+4.82}_{-4.08}$&$61.00^{+1.80}_{-1.62}$&$60.82^{+0.90}_{-0.86}$\\
$\sigma_{(|(c\bar{c})[n^3P_2]\rangle)}$&$30.18^{+1.50}_{-1.74}$&$14.98^{+1.61}_{-1.34}$&$20.43^{+0.59}_{-0.86}$&$20.37^{+0.29}_{-0.28}$\\
\hline
Sum~~&$549.0^{+16.2}_{-36.8}$&$288.3^{+10.6}_{-8.7}$&$217.0^{+4.2}_{-3.5}$&$189.2^{+13.5}_{-12.3}$\\
\hline\hline
$\sigma_{(|(b\bar{b})[n^1S_0]\rangle)}$&$52.76^{+1.82}_{-1.82}$&$20.73^{+0.95}_{-0.96}$&$6.462^{+0.262}_{-0.214}$&$7.550^{+0.025}_{-0.040}$\\
$\sigma_{(|(b\bar{b})[n^3P_0]\rangle)}$&$0.715^{+0.006}_{-0.005}$&$0.308^{+0.013}_{-0.012}$&$0.267^{+0.013}_{-0.013}$&$0.302^{+0.009}_{-0.009}$\\
$\sigma_{(|(b\bar{b})[n^3P_1]\rangle)}$&$4.664^{+0.001}_{-0.008}$&$2.019^{+0.098}_{-0.095}$&$1.759^{+0.073}_{-0.071}$&$1.997^{+0.148}_{-0.046}$\\
$\sigma_{(|(b\bar{b})[n^3P_2]\rangle)}$&$1.592^{+0.003}_{-0.003}$&$0.691^{+0.035}_{-0.034}$&$0.602^{+0.024}_{-0.023}$&$0.685^{+0.015}_{-0.014}$\\
\hline
Sum~~&$59.73^{+1.83}_{-1.84}$&$23.75^{+1.10}_{-1.10}$&$9.091^{+0.372}_{-0.321}$&$10.53^{+0.20}_{-0.11}$\\
\hline\hline
\end{tabular}
\label{amUncertanty}
\end{table}
\begin{table}
\caption{Uncertainties of total cross sections (units: $\times 10^{-2}$fb) caused by varying the masses as shown in Table \ref{M&R} for $Z^0$ propagated processes. Note, effects of uncertainties of radial wave functions at the origin and their first derivatives at the origin caused by varying masses are also considered.}
\begin{tabular}{|c|c|c|c|c|}
\hline
$|(Q\bar{Q})[n]\rangle$&~$n=1$~&~$n=2$~&~$n=3$~&~$n=4$~\\
\hline\hline
$\sigma_{(|(c\bar{c})[n^1S_0]\rangle)}$&$239.8^{+5.5}_{-16.8}$&$128.2^{+1.9}_{-1.5}$&$72.82^{+0.88}_{-0.41}$&${56.86}^{+1.64}_{-1.11}$\\
$\sigma_{(|(c\bar{c})[n^3S_1]\rangle)}$&$1632^{+38}_{-115}$&$873.2^{+13.4}_{-10.4}$&$496.0^{+6.1}_{-2.9}$&$387.3^{+11.4}_{-7.5}$\\
$\sigma_{(|(c\bar{c})[n^1P_1]\rangle)}$&$177.1^{+9.0}_{-9.8}$&$87.71^{+9.70}_{-7.85}$&$119.8^{+3.6}_{-3.3}$&${119.3}^{+1.8}_{-1.7}$\\
$\sigma_{(|(c\bar{c})[n^3P_0]\rangle)}$&$8.620^{+0.429}_{-0.472}$&$4.261^{+0.475}_{-0.385}$&$5.807^{+0.178}_{-0.164}$&$5.776^{+0.094}_{-0.091}$\\
$\sigma_{(|(c\bar{c})[n^3P_1]\rangle)}$&$52.26^{+2.67}_{-2.92}$&$25.89^{+2.85}_{-2.31}$&$35.38^{+1.03}_{-0.95}$&${35.26}^{+0.52}_{-0.50}$\\
$\sigma_{(|(c\bar{c})[n^3P_2]\rangle)}$&$17.47^{+0.90}_{-0.98}$&$8.664^{+0.951}_{-0.769}$&$11.84^{+0.34}_{-0.31}$&$11.81^{+0.17}_{-0.16}$\\
\hline
Sum~~&$2128^{+56}_{-146}$&$1128^{+29}_{-23}$&$741.6^{+12.1}_{-8.0}$&$616.3^{+15.6}_{-11.1}$\\
\hline\hline
$\sigma_{(|(b\bar{b})[n^1S_0]\rangle)}$&$398.1^{+13.7}_{-13.7}$&$156.4^{+7.2}_{-7.4}$&$48.76^{+1.98}_{-1.61}$&$56.96^{+0.19}_{-0.30}$\\
$\sigma_{(|(b\bar{b})[n^3S_1]\rangle)}$&$840.8^{+29.8}_{-29.7}$&$330.8^{+15.5}_{-15.9}$&$103.2^{+4.3}_{-3.5}$&$120.6^{+0.5}_{-0.8}$\\
$\sigma_{(|(b\bar{b})[n^1P_1]\rangle)}$&$35.91^{+0.07}_{-0.08}$&$15.52^{+0.72}_{-0.70}$&$13.51^{+0.59}_{-0.57}$&$ 15.31^{+0.40}_{-0.38}$\\
$\sigma_{(|(b\bar{b})[n^3P_0]\rangle)}$&$5.395^{+0.037}_{-0.039}$&$2.322^{+0.095}_{-0.094}$&$2.017^{+0.102}_{-0.096}$&$2.275^{+0.072}_{-0.069}$\\
$\sigma_{(|(b\bar{b})[n^3P_1]\rangle)}$&$35.19^{+0.01}_{-0.01}$&$15.24^{+0.74}_{-0.72}$&$13.27^{+0.55}_{-0.54}$&$15.07^{+0.36}_{-0.35}$\\
$\sigma_{(|(b\bar{b})[n^3P_2]\rangle)}$&$12.01^{+0.02}_{-0.02}$&$ 5.210^{+0.265}_{-0.255}$&$4.543^{+0.180}_{-0.174}$&$5.165^{+0.114}_{-0.108}$\\
\hline
Sum~~&$1328^{+44}_{-43}$&$525.5^{+24.5}_{-25.1}$&$185.3^{+7.7}_{-6.5}$&$215.4^{+1.6}_{-2.0}$\\
\hline\hline
\end{tabular}
\label{ZmUncertanty}
\end{table}

For the leading-order calculation, the main uncertainty sources of cross sections include the Fermi constant $G_F$, the Weinberg angle $\theta_W$, the fine-structure constant $\alpha$, the mass and width of  the $Z^0$ boson, the masses of constituent quarks, and the non-perturbative matrix elements. 
Since parameters $G_F$, $\theta_W$, $\alpha$ and the mass and width of  the $Z^0$ boson are either an overall factor or an relatively precise value, we will not discuss uncertainties caused by them. 
In this subsection, we will explore uncertainties caused by masses of constituent quarks, the non-perturbative matrix elements, and deviation of CM energy $\sqrt{s}$ away from $m_Z$. 

The uncertainties of cross sections caused by varying the masses of constituent quarks by 0.1 GeV for $m_c$ and 0.2 GeV for $m_b$ (as shown in Table \ref{M&R} ) at the CM energy $\sqrt{s}=91.1876$ GeV are presented in Tables~~\ref{amUncertanty} and \ref{ZmUncertanty} for virtual photon $\gamma^*$ and $Z^0$ propagated processes, respectively.
It worth noting that the effects of uncertainties of radial wave functions at the origin and their first derivatives at the origin caused by varying masses are also taken into consideration.
It is found that the wave functions at the origin and their derivatives at the origin increase as quark masses increase. 
But, we find that the short-distance coefficients decrease along with the increasement of quark masses. 
The overall effect is that the cross sections decrease with the increment of the quark masses.

We adopt four other potential models to estimate the uncertainties caused by the wave functions at the origin and their first derivatives at the origin in Tables \ref{wavefunc1} and \ref{wavefunc2} for $Z^0$ propagated processes for charmonium and bottomonium, respectively. 
The four models are QCD-motivated potential with one-loop correction given by John L. Richardson (J. potential) \cite{Richardson:1978bt}, 
QCD-motivated potential with two-loop correction given by K. Igi and S. Ono (I.O. potential) \cite{IO,Ikhdair:2003ry}, QCD-motivated potential with two-loop correction given by Yu-Qi Chen and Yu-Ping Kuang (C.K. potential) \cite{Chen:1992fq,Ikhdair:2003ry}, and the QCD-motivated Coulomb-plus-linear potential (Cor. potential) \cite{Ikhdair:2003ry,Eichten:1978tg,Eichten:1979ms,Eichten:1980mw,Eichten:1995ch}. 
The formula and latest values of those wave functions at the origin and their first derivatives at the origin can be found in our earlier work \cite{lx}. 
In Tables \ref{wavefunc1} and \ref{wavefunc2}, the contributions from four P-wave states ($n^1P_1$, $n^3P_J$ with $J=0,1,2$) are summed up.
It is shown that the cross sections change dramatically when we choose different potential models. 
For the production of $|(c\bar{c})[n]\rangle$ in Table \ref{wavefunc1}, we always obtain the minimum under the I.O. potential model, and obtain the maximum under the B.T. potential or J. potential models. 
While for the production of $|(b\bar{b})[n]\rangle$ in Table \ref{wavefunc2}, we obtain the minimum under C.K. or I.O. potential models, and obtain the maximum under the B.T., or Cor. potential models. 
In Tables \ref{wavefunc1} and \ref{wavefunc2}, percentages in brackets are the ratios of the minimum or maximum relative to the estimates under the B.T. model.
\begin{table*}
\caption{Uncertainties of total cross sections (units: $\times 10^{-2}$ fb) caused by five different potential models for $|(c\bar{c})[n]\rangle$ quarkonium in $e^+e^-\to Z^0\to |(c\bar{c})[n]\rangle+\gamma$. Percentages in brackets are the ratios of the minimum or maximum relative to the estimates under B.T. model.}
\begin{tabular}{|c|c|c|c|c|c|}
\hline
$|(c\bar{c})[n]\rangle$&~B.T.~&~J.~&~I.O.~&~C.K.~&~Cor.~\\
\hline\hline
$\sigma_{(|(c\bar{c})[1^1S_0]\rangle)}$&~239.8~&~109.2~&~55.12(23\%)~&~70.82~&~95.02~\\
\hline
$\sigma_{(|(c\bar{c})[1^3S_1]\rangle)}$&~1632~&~743.1~&~375.2(23\%)~&~482.1~&~646.8~\\
\hline
$\sigma_{(|(c\bar{c})[1P]\rangle)}$&~255.5~&~136.5~&~42.05(16\%)~&~58.71~&~72.20~\\
\hline\hline
$\sigma_{(|(c\bar{c})[2^1S_0]\rangle)}$&~128.2~&~83.81~&~43.53(34\%)~&~48.68~&~70.48~\\
\hline
$\sigma_{(|(c\bar{c})[2^3S_1]\rangle)}$&~873.2~&~570.8~&~296.5(34\%)~&~331.6~&~480.0~\\
\hline
$\sigma_{(|(c\bar{c})[2P]\rangle)}$&~126.5~&~174.5(138\%)~&~55.92(44\%)~&~72.30~&~95.46~\\
\hline\hline
$\sigma_{(|(c\bar{c})[3^1S_0]\rangle)}$&~72.82~&~74.02(102\%)~&~38.93(53\%)~&~41.93~&~61.69~\\
\hline
$\sigma_{(|(c\bar{c})[3^3S_1]\rangle)}$&~496.0~&~504.2(102\%)~&~265.1(53\%)~&~285.6~&~420.2~\\
\hline
$\sigma_{(|(c\bar{c})[3P]\rangle)}$&~172.8~&~195.(113\%)~&~83.04(48\%)~&~100.9~&~142.9~\\
\hline\hline
$\sigma_{(|(c\bar{c})[4^1S_0]\rangle)}$&~56.86~&~69.29(122\%)~&~36.65(64\%)~&~38.72~&~57.65~\\
\hline
$\sigma_{(|(c\bar{c})[4^3S_1]\rangle)}$&~387.3~&~471.9(122\%)~&~249.6(64\%)~&~263.7~&~392.6~\\
\hline
$\sigma_{(|(c\bar{c})[4P]\rangle)}$&~172.1~&~208.6(121\%)~&~68.55(40\%)~&~83.30~&~117.9~\\
\hline\hline
Sum~~&~4613~&~3341~&~1610(35\%)~&~1878~&~2653~\\
\hline
\end{tabular}
\label{wavefunc1}
\end{table*}
\begin{table*}
\caption{Uncertainties of total cross sections (units: $\times 10^{-2}$ fb) caused by five different potential models for $|(b\bar{b})[n]\rangle$ quarkonium in $e^+e^-\to Z^0\to |(b\bar{b})[n]\rangle+\gamma$. Percentages in brackets are the ratios of the minimum or maximum relative to the estimates under B.T. model.}
\begin{tabular}{|c|c|c|c|c|c|}
\hline
$|(b\bar{b})[n]\rangle$&~B.T.~&~J.~&~I.O.~&~C.K.~&~Cor.~\\
\hline\hline
$\sigma_{(|(b\bar{b})[1^1S_0]\rangle)}$&~398.1~&~175.7~&~246.5~&~130.8(33\%)~&~225.7~\\
\hline
$\sigma_{(|(b\bar{b})[1^3S_1]\rangle)}$&~840.8~&~371.1~&~520.6~&~276.3(33\%)~&~476.7~\\
\hline
$\sigma_{(|(b\bar{b})[1P]\rangle)}$&~88.50~&~24.77~&~17.55~&~16.74(19\%)~&~18.35~\\
\hline\hline
$\sigma_{(|(b\bar{b})[2^1S_0]\rangle)}$&~156.4~&~96.12~&~80.26~&~64.52(41\%)~&~110.6~\\
\hline
$\sigma_{(|(b\bar{b})[2^3S_1]\rangle)}$&~330.8~&~203.3~&~169.8~&~136.5(41\%)~&~233.9~\\
\hline
$\sigma_{(|(b\bar{b})[2P]\rangle)}$&~47.38~&~35.97~&~16.17(34\%)~&~22.19~&~27.97~\\
\hline\hline
$\sigma_{(|(b\bar{b})[3^1S_0]\rangle)}$&~48.76~&~76.35~&~46.04(94\%)~&~49.83~&~87.57(180\%)~\\
\hline
$\sigma_{(|(c\bar{c})[3^3S_1]\rangle)}$&~103.2~&~161.6~&~97.45(94\%)~&~105.5~&~185.4(180\%)~\\
\hline
$\sigma_{(|(b\bar{b})[3P]\rangle)}$&~33.35~&~31.73~&~10.27(31\%)~&~18.76~&~25.37~\\
\hline\hline
$\sigma_{(|(b\bar{b})[4^1S_0]\rangle)}$&~56.96~&~67.07~&~32.00(56\%)~&~43.41~&~77.02(135\%)~\\
\hline
$\sigma_{(|(b\bar{b})[4^3S_1]\rangle)}$&~120.6~&~142.0~&~67.78(56\%)~&~91.92~&~163.1(135\%)~\\
\hline
$\sigma_{(|(b\bar{b})[4P]\rangle)}$&~27.61~&~24.38~&~6.008(22\%)~&~14.04~&~19.95~\\
\hline\hline
Sum~~&~2252~&~1410~&~1310~&~970.5(43\%)~&~1652~\\
\hline
\end{tabular}
\label{wavefunc2}
\end{table*}

For the uncertainties of total cross sections caused by the deviation of CM energy $\sqrt{s}$ away from $m_Z$, one can have a visual impression in Figs. \ref{ccds} and \ref{bbds}. 
It is shown that the cross sections decreases dramatically with the deviation of CM energy $\sqrt{s}$ away from $m_Z$. To obtain a quantitative impression, we display the uncertainties caused by the deviation of CM energy $\sqrt{s}$ away from $m_Z$ by 1\% and 3\% for the $Z^0$ propagated process with $n=1$ in Table \ref{sqrtsUncertanty}.
\begin{table}
\caption{Uncertainties of total cross sections (units: $\times 10^{-2}$ fb) caused by the deviation of CM energy $\sqrt{s}$ away from $m_Z$ for $|(Q\bar{Q})[1]\rangle$ quarkonium in $e^+e^-\to Z^0 \to |(Q\bar{Q})[1]\rangle+\gamma$ under B.T. model.}
\begin{tabular}{|c|c|c|c|c|c|}
\hline
~$\sqrt{s}$~&$97\%m_Z$&$99\%m_Z$&$m_Z$&$101\%m_Z$&$103\%m_Z$\\
\hline\hline
$\sigma(|(c\bar{c})[1^1S_0]\rangle)$&~42.33~&~156.8~&~239.8~&~155.8~&~40.28~\\
\hline
$\sigma(|(c\bar{c})[1^3S_1]\rangle)$&~288.1~&~1068~&~1632~&~1060~&~274.2~\\
\hline
$\sigma{(|(c\bar{c})[1^1P_1]\rangle)}$&~31.27~&~115.9~&~177.1~&~115.1~&~29.75~\\
\hline
$\sigma{(|(c\bar{c})[1^3P_0]\rangle)}$&~1.521~&~5.637~&~8.620~&~5.597~&~1.448~\\
\hline
$\sigma{(|(c\bar{c})[1^3P_1]\rangle)}$&~9.227~&~34.19~&~52.26~&~33.95~&~8.778~\\
\hline
$\sigma{(|(c\bar{c})[1^3P_2]\rangle)}$&~3.084~&~11.43~&~17.47~&~11.35~&~2.934~\\
\hline
Sum~~&~375.6~&~1392~&~2128~&~1382~&~357.4~\\
\hline\hline
$\sigma(|(b\bar{b})[1^1S_0]\rangle)$&~70.23~&~260.3~&~398.1~&~258.6~&~66.91~\\
\hline
$\sigma(|(b\bar{b})[1^3S_1]\rangle)$&~148.4~&~549.9~&~840.8~&~546.1~&~141.2~\\
\hline
$\sigma{(|(b\bar{b})[1^1P_1]\rangle)}$&~6.339~&~23.48~&~35.91~&~23.32~&~6.032~\\
\hline
$\sigma{(|(b\bar{b})[1^3P_0]\rangle)}$&~0.949~&~3.517~&~5.395~&~3.502~&~0.908~\\
\hline
$\sigma{(|(b\bar{b})[1^3P_1]\rangle)}$&~6.221~&~23.06~&~35.19~&~22.88~&~5.911~\\
\hline
$\sigma{(|(b\bar{b})[1^3P_2]\rangle)}$&~2.127~&~7.883~&~12.01~&~7.813~&~2.017~\\
\hline
Sum~~&~234.3~&~868.2~&~1328~&~862.3~&~223.0~\\
\hline
\end{tabular}
\label{sqrtsUncertanty}
\end{table}

\section{Conclusions}
In the present work, we make a comprehensive study on the high excited states of the $|(c\bar{c})[n]\rangle$ and $|(b\bar{b})[n]\rangle$ quarkonium production in $e^+e^-\to \gamma^*/Z^0\to |(Q\bar{Q})[n]\rangle+\gamma$ within the NRQCD factorization framework at  future $Z$ factory, where $[n]$ stands for $[n^1S_0]$, $[n^3S_1]$, $[n^1P_1]$, and $[n^3P_J]$ Fock states ($n=1, 2, 3, 4$; $J=0, 1, 2$). 
The``improved trace technology", which disposes the Dirac matrices at the amplitude level, is helpful for deriving compact analytical results especially for the complicated $P$-wave processes with massive spinors. 
The total cross sections $\sigma(\sqrt{s})$ and differential distributions $d\sigma/dcos\theta$ and $d\sigma/dp_{t}$ for all $n=1$ Fock states are studied in detail. 
For a sound estimation, we further study the uncertainties of the cross sections caused by the varying mass of $c$ and $b$ quarks, the non-perturbative matrix elements under five potential models, and deviation of CM energy $\sqrt{s}$ away from $m_Z$.

In addition to the ground states, it is found that the production rates of high excited Fock states of charmonium and bottomonium are considerable in the processes of $e^+e^-\to \gamma^*/Z^0\to |(Q\bar{Q})[n]\rangle+\gamma$ at super $Z$ factory with high luminosity ${\cal L}\approx 10^{36}cm^{-2}s^{-1}$. 
The cross sections of charmonium for $2S$, $3S$, $4S$, $1P$, $2P$, $3P$ and $4P$-wave states are about $53.5\%$, $30.4\%$, $23.7\%$, $13.7\%$, $6.8\%$, $9.2\%$, and $9.2\%$ of that of the $1S$ state, respectively. 
And cross sections of bottomonium for $2S$, $3S$, $4S$, $1P$, $2P$, $3P$ and $4P$-wave states are about $39.3\%$, $12.3\%$, $14.3\%$, $7.1\%$, $3.1\%$, $2.7\%$, and $3.1\%$ of that of the $1S$ state, respectively. 
Then, such a super $Z$ factory could provide a useful platform to study the high excited charmonium and bottomonium.
In addition, we find that cross sections change dramatically when adopting different potential models, which would be the major source of uncertainty. 
And the deviation of CM energy $\sqrt{s}$ away from $Z^0$ pole at future super $Z$ factory will also have great influence on the production rates. 

\hspace{2cm}

{\bf Acknowledgements}: 
This work was supported in part by 
the National Natural Science Foundation of China under Grant No. 11905112, 
and the Natural Science Foundation of Shandong Province under Grant No. ZR2019QA012.

\end{document}